\documentstyle[12pt]{article}
\title{Numerical Study of $ K^0 - \bar{K}^0 $ Mixing and
$ B_K $}
\author{
Weonjong Lee\thanks{Current address: IBM Thomas J. Watson Research Center,
Room 27-161, P.O. Box 218, Yorktown Heights, NY 10598}
and
Markus Klomfass\thanks{Current address: Veilchenweg 24,
65201 Wiesbaden, Germany} \\
{\small \it Department of Physics,
Pupin Physics Laboratories,}\\
{\small \it Columbia University,
New York, N.Y. 10027}
}
\date{\today}
\begin{document}
\maketitle
\begin{abstract}
We have computed $ B_K $ 
with staggered fermions,
using two different methods.
The numerical simulations were performed
on a $ 16^3 \times 40 $ lattice in full QCD
with $ \beta = 5.7 $.
We also tried an improved wall source method
in order to select only the pseudo-Goldstone bosons and
compare the numerical results obtained with
those from the conventional wall source method.
We have studied a series of non-degenerate quark anti-quark pairs
and saw no effect on $ B_K $,
although effects were seen on the individual
terms making up $ B_K $.
\end{abstract}
\section{Introduction}
In the standard model,
there are two kinds of CP violation: the indirect
CP violation and the direct CP violation.
\cite{buras10,jarlskog,wise1,paschos}. 
The indirect CP violation
comes from the fact that electro-weak
$ K^0 $-$ \bar{K}^0 $ mixing causes
the neutral kaon eigenstates not to
respect CP symmetry \cite{b-lee}. 
The direct CP violation
comes from the CP violating effective
operators ({\em e.g.} penguin operators)
\cite{wise1,paschos}, which can directly contibute to
the K decay amplitude.
The indirect (direct) CP violation is parametrized
by the phenomenological quantity $ \varepsilon $,
while the direct CP violation is parametrized
by $ \varepsilon' $ \cite{buras10,jarlskog,wise1,paschos}.
Both $ \varepsilon $ and  $ \varepsilon' $ can be 
measured experimentally.
The low energy effective
Hamiltonian of the electro-weak interaction
is derived by decoupling heavy particles
such as $ W $ and $ Z $ bosons and the t, b, c
quarks in the standard model
\cite{wise1,wise2,wise3,buras20}.
The coefficients in this effective Hamiltonian
are functions of the Cabibbo-Kobayashi-Maskawa
flavor mixing matrix elements $ V_{CKM} $.
This low energy effective Hamiltonian
includes a $ \Delta S = 2 $ term
which belongs to the (27,1) representation 
of the $ SU(3)_L \otimes SU(3)_R $
flavor symmetry group and to the (1,0) representation
of the $ SU(2)_L \otimes SU(2)_R $ isospin symmetry subgroup. 
This $ \Delta S = 2 $ electro-weak
effective Hamiltonian 
causes the $ K^0 $-$ \bar{K}^0 $ mixing
in the standard model
\cite{wise1,wise2,wise3,buras20}. 
Therefore, the $ \Delta S = 2 $ effective Hamiltonian
is connected to the $ \varepsilon $ parameter.
Using this technique, $ \varepsilon $ can be expressed in terms 
of the standard model parameters
as follows:
\begin{eqnarray}
\varepsilon = C \hat{B}_K \ F(V_{CKM}, \frac{m_t^2}{M_W^2},
\frac{m_c^2}{M_W^2}, \frac{m_u^2}{M_W^2}),
\end{eqnarray}
where
\begin{eqnarray}
& & C = \frac{G_F^2 f_K^2 m_K M_W^2}{6\sqrt{2}\pi^2[ m(K_L) -m(K_S) ]}
\\
& & B_K = \frac{
\langle K^0 \mid [\bar{s} \gamma_{\mu}(1 - \gamma_5) d]
[\bar{s} \gamma_{\mu}(1 - \gamma_5) d] \mid K'^0 \rangle }
{ \frac{8}{3}
\langle K^0 \mid
\bar{s} \gamma_{\mu} \gamma_5 d \mid 0 \rangle
\langle 0 \mid
\bar{s} \gamma_{\mu} \gamma_5 d \mid K'^0 \rangle }
\\
& & \hat{B}_K =
\xi(\alpha_s(\mu))  B_K(\mu) \ .
\end{eqnarray}
Here $ \hat{B}_K $ is defined as a 
renormalization-group invariant quantity,
and the function \linebreak
$ F(V_{CKM}, m_t^2/M_W^2, \cdots) $
is given in Ref. \cite{buras10,wise1,paschos}.
Once $ B_K $ is determined theoretically,
we can narrow the domain of $ \mid V_{td} \mid $ and
the top quark mass \cite{buras10},
using the experimental determination of $ \varepsilon $.
%


%
There have been a variety of methods used to calculate
$ B_K $, including chiral perturbation theory, 
hadronic sum rules, QCD sum rules, $ 1/N_C $ expansion
and lattice gauge theory.
Lattice gauge theory has the virtue
of making the smallest number of assumptions and is 
exactly equivalent to QCD in the limit of infinite volume
and vanishing lattice spacing.
For this reason, we have adopted lattice gauge theory to
obtain $ B_K $.
In order to calculate $ B_K $ 
in lattice gauge theory,
one needs to find a set of operators
which can describe on the lattice the same physics
as the continuum $ \Delta S = 2 $ four-fermion operator.
There have been two methods to implement 
fermions in lattice gauge theory:
the staggered fermion method and Wilson fermion
method.
For the weak matrix elements involving
the pseudo-Goldstone bosons, it is very useful
to take advantage of the exact $ U_A(1) $ 
symmetry of the staggered fermion action,
which is not manifest
in Wilson fermion action \cite{sharpe0,sharpe1}.
We have used the staggered fermion method to
calculate $ B_K $ because of this advantage.
There are two methods to transcribe the continuum
$ \Delta S =2 $ weak matrix elements to the lattice 
with staggered fermions \cite{sharpe0,wlee0}:
the one spin trace formalism and 
the two spin trace formalism.
The four fermion operators can be expressed as products
of operators bilinear in the fermion fields.
In the one spin trace formalism, 
each external hadron is contracted
with both bilinears of the four fermion operator
simultaneously.
In the two spin trace formalism, each external hadron is
contracted with only one of the bilinears in the four
fermion operator.
Until now, the two spin trace formalism (2TR) has
been used exclusively in calculations of
weak matrix elements with staggered fermions,
\cite{sharpe19,sharpe20,sharpe21},
\cite{kilcup10,kilcup0,kilcup20,kilcup30}
and \cite{japan0,japan20}.
Recently, the one spin trace formalism has been
developed to a level which permits it to be used
for numerical simulations of weak matrix elements
such as $ B_K $
\cite{wlee0}.
We have used both formalisms to calculate $ B_K $
and the results are compared in this paper.
Lattice calculations introduce their own
systematic artifacts and errors which can be sizable.
One of the principal sources of systematic errors
is finite volume.
The results of a finite volume comparison
were reported in Ref. \cite{kilcup0}, where 
it was argued that the finite volume effects
on $ B_K $ are quite small
when $ V \times T \ge 16^3 \times 40 $ 
at $ \beta = 6.0 $ ($a^{-1} \sim 2$ GeV),
$ N_f = 0 $.
Another source of systematic errors comes from
the quenched approximation,
neglecting all of the internal quark loops.
Quenched effects were studied in Ref. \cite{kilcup10},
which concluded that the effect of quenching
can not be large.
Another systematic error which is called {\em scaling violation}
comes from the finite lattice spacing ($ a \neq 0 $). 
In Ref. \cite{sharpe21}, the scaling corrections
were argued to be of $ a^2 $ order. 
In other words,
$ O(a) $ corrections do not exist 
at all for the staggered
fermion operators of $ B_K $.
Related to possible scaling violations is a possible
discrepancy between gauge-dependent Landau gauge operators
and  gauge invariant operators.
The Landau gauge operator implies 
that the quark anti-quark propagators are 
fixed to Landau gauge and that gauge links between the
quark and anti-quark fields are omitted. 
The question was raised as to whether the
Landau gauge operators might cause
the large scaling violation  which had been
noticed originally in Ref. \cite{sharpe19}.
However, it was reported in Ref. \cite{japan0}
that the results of both Landau gauge
operators and the gauge invariant 
operators were numerically
found to be consistent with each other.
The purpose of this paper is to report and analyze the numerical data
for $ B_K $ as well as its individual terms and to interpret
the numerical results in terms of various physical
models.
Part of the preliminary results
have already appeared in Ref. \cite{wlee10}. 
In this paper we will address the following five issues
through the interpretation of our numerical results.
The first issue is how to select the pseudo
Goldstone boson state exclusively. 
For  hadron spectrum measurements, the sink
operator picks up a specific hadronic state
exclusively.
In contrast to hadron spectrum measurements,
the operators for weak matrix element measurements
do not select a particular hadronic state.
We need to impose a symmetry requirement
on the wall source such that all the unwanted
states are decoupled in the weak matrix element 
measurement. 
We have tried an improved wall source method
(called {\em cubic wall source} \cite{japan10}) to do this.
The second issue is whether the numerical
results in the one spin trace form are
in agreement with those in the two spin trace form.
Theoretically
the difference between the two formalisms
is supposed to vanish in the limit of 
$ a \rightarrow 0 $ \cite{wlee0} for $ B_K $.
We have tried both the one spin trace formalism
and the two spin trace formalism to calculate
$ B_K $. The results are compared in this paper.
The third issue is the effect of non-degenerate
quark antiquark pairs on $ B_K $ and the individual
components making up $ B_K $.
The kaon is composed of $ s $ and $ d $ valence quarks.
Here a non-degenerate quark anti-quark pair implies
that the $ s $ valence quark mass is different from 
the $ d $ valence quark mass, whereas a degenerate
quark anti-quark pair implies
that both valence flavors have the
same mass.
The effect of non-degenerate valence quarks 
on $ B_K $ in quenched QCD was mentioned briefly in Ref.
\cite{sharpe20} where a small difference of only marginal
significance was found. 
In this paper, we investigate in detail
the effects of non-degenerate valence quark 
anti-quark pairs on $ B_K $ and its individual components.
From the theoretical point of view,
one effect of non-degenerate valence quark
anti-quark pairs
can be related to the $ \eta' $ hairpin diagram
in (partially) quenched QCD
\cite{sharpe30,bernard0,bernard1}.
The forth issue is whether quenched
chiral perturbation theory is compatible with
the numerical results of $ B_K $ \cite{sharpe30}.
Quenched chiral perturbation theory
also predicts the chiral behavior of the individual
terms making up $ B_K $ \cite{sharpe30}.
It is good to know how reliable these theoretical
predictions are numerically.
The final issue is whether there is any dynamical fermion
effect on $ B_K $.
This question was addressed originally
in Ref. \cite{kilcup10}.
We will re-visit this question and see how important
the internal fermion loops are to $ B_K $.
It is important to understand the difference
between full QCD and quenched QCD in $ B_K $
both theoretically and numerically. 
This paper is organized as follows.
In section 2, we will describe the technical 
details in doubling the lattice for quark propagators
and explain the improved (cubic) wall source method in a
self-contained manner.
In Section 3, we will specify the lattice operators
for $ B_K $ in brief while leaving the details
to adequate references.
In Section 4, we describe the parameters for
the gauge configurations we generated and the measurement
parameters for $ B_K $.
In Section 5, we present the numerical data
for  $ {\cal M}_K $ (numerator of $ B_K $)
and the vacuum saturation amplitude (denominator of $ B_K $).
The main emphasis is put on the determination
of the plateau region.  
We also discuss the improved wall source
with the wrong flavor channel, which is supposed to vanish
in the limit of $ a \rightarrow 0 $.
In Section 6, the numerical results from the data analysis
are interpreted. 
The improved (cubic) wall source method is compared with
the conventional method. 
The data of the one spin trace form
and of the two spin trace form are compared. 
Enhanced chiral logarithms in the individual
terms making up $ B_K $ appears to be seen numerically
as well as the additional divergence
which arises from the effect of the non-degenerate 
quark anti-quark pairs.
We emphasize the large effect of
non-degenerate quark anti-quark pairs on the individual
components of $ B_K $.
We compare our data with other groups (Kilcup,
Sharpe and Ukawa {\em et al.}).
We describe the covariance fitting procedure
for $ B_K $ and the negligible effect of
non-degenerate quark anti-quark pair on $ B_K $.
Section 7 contains a brief summary and our conclusions.
\section{Quark Propagators and Wall Sources}
Here, we explain the technical details 
of the quark propagators and the improved wall source
method.
Because the lattice size in the time direction is finite,
the pions propagating around lattice in the time direction 
can in principle contaminate the measurements of $ B_K $.
There have been two proposals for avoiding this contamination.
The first proposal is to impose 
Dirichlet boundary conditions in time
on the quark propagator 
and to place the wall source near the boundary 
\cite{sharpe1,kilcup10,japan0}. 
The second proposal is to double the lattice
along the time direction and to use periodic
boundary conditions in time for the quark propagator
\cite{sharpe1,sharpe20}.
Dirichlet boundary conditions
cause a certain number of time slices near the wall source
to be contaminated by reflections off the boundary.
The time slices lost due to reflections
overlap with those contaminated by $ \rho $
mesons.
Lattice doubling in the time direction suppresses 
the backward propagating pions 
which must travel over the whole lattice size in the time
direction before they can contribute to the measurements. 
In contrast to Dirichlet boundary conditions,
lattice doubling does not introduce any reflection from the
boundary.
Hence, the signal from lattice doubling
is much cleaner than that obtained using 
Dirichlet boundary conditions.
Unfortunately, lattice doubling takes twice the
computational time since the lattice doubling
needs two undoubled quark propagators 
for a given wall source (forward and backward) 
instead of the one quark propagator
required when using Dirichlet boundary conditions.
We adopt the second solution of doubling the
lattice for our numerical simulation of 
$ B_K $. 
Here, we describe this second method in detail.
The periodic and anti-periodic
quark propagators in time on the undoubled lattice
are:
\begin{eqnarray}
& & \sum_{x}(D + M )_{(z,x)} \cdot G_P(x,y) = \delta_{z,y} \\
\mbox{and } & & \nonumber \\
& & \sum_{x}(D + M )_{(z,x)} \cdot G_A(x,y) = \delta_{z,y} \  ,
\end{eqnarray}
where $ (D + M ) $ represents the Euclidean
Dirac operator, and $ G_P(x,y) $ and $ G_A(x,y) $
represent the Green's 
functions with periodic and  anti-periodic
boundary conditions on the undoubled lattice.
The source $h(y)$ is introduced in the numerical simulation as
follows:
for $ 0 \leq x_t < N_t $,
\begin{eqnarray}
& & \chi(x) = \frac{1}{2} 
\sum_{y} ( G_P(x,y) + G_A(x,y) ) \cdot h(y) \ , \\
& & \chi(x+N_t \hat{t}) = \frac{1}{2}
\sum_{y} ( G_P(x,y) - G_A(x,y) ) \cdot h(y)
\end{eqnarray}
where $ \hat{t} $ is the unit vector
in time direction and $ 0 \leq y_t < N_t $
($ N_t $ is the undoubled lattice size in the time direction). 
The $ \chi(x) $ field in the above satisfies the following:
for $ 0 \leq x_t < N_t $,
\begin{eqnarray}
\sum_x (D + M)_{(z,x)} \cdot \chi(x) = h(z)
\end{eqnarray}
and for $ N_t \le x_t < 2N_t$,
\begin{eqnarray}
\sum_x (D + M)_{(z,x)} \cdot \chi(x + N_t \hat{t}) = 0 \ .
\end{eqnarray}
We use periodic and anti-periodic 
boundary conditions on the undoubled lattice 
and take the sum and the difference
to obtain a quark propagator
which is periodic on the doubled lattice 
in the time direction.
Note that the fermionic field
$ \chi(x) $ is periodic on the doubled lattice
in the time direction.
One might ask whether  periodic boundary conditions
in the time direction make any difference compared to 
the anti-periodic boundary conditions, which are the
physical ones at a finite temperature.
The answer is that periodic boundary conditions
make no difference 
as long as the volume is large enough. 
The physical eigenstates in the confined phase of QCD
are hadrons, not quarks nor gluons.
In the confined phase, a quark must 
be confined with other quarks or
anti-quarks within a small volume of
typical hadronic size ($ \approx $ $ {\cal O}( 1 fm) $)
before it can acknowledge the existence of the boundary 
with the volume much larger than the hadronic size. 
There have been a number of attempts to enhance the overlap
with the lightest particle compared to that of the
excited states so that one can see the asymptotic
signal ($ \exp( -M \mid t \mid) $) at smaller time
separations and over the longer plateau
\cite{kilcup10,japan10,gupta0}.
The wall source method has an advantage because it enhances
the signal of the hadron propagators with respect to
the point source method.
For hadron mass spectrum measurements,
the sink operator possesses the symmetry
of a specific hadronic state.
In contrast to
hadron spectrum measurements,
since the operator in the electro-weak effective Hamiltonian
does not select any particular hadronic state
by itself, the weak matrix
element measurements require the symmetry
properties of the wall source to determine
the specific hadronic state.
There have been two attempts to improve the wall
source such that it can exclude  contamination
from  unwanted hadronic states:
the even-odd wall source method
\cite{kilcup20,gupta0} and 
the cubic wall source method originally
proposed by M. Fukugita, 
{\em et al.} in Ref. \cite{japan10}.
Here, we explain 
the cubic wall source method in detail.
The $ B_K $ measurements requires that
the pseudo-Goldstone Kaon mode should be selected exclusively.
Hence, we will restrict our discussion
to the cubic wall source operator
of pseudo-Goldstone mode.
Let us start with definitions and
notation.
The symbol $ \bf \bar{a} $ represents one of the 8 vertices
in the unit spatial cube
({\em i.e.} $ {\bf \bar{a}} \in \{0,1\}^3 $): 
\begin{eqnarray}
{\bf \bar{a}} \in 
\left\{ (0,0,0), (1,0,0), (0,1,0), \cdots, (1,1,1) \right\} \ . 
\end{eqnarray}
We define $ W_{ \bf \bar{a} }(\vec{n}) $ as follows:
\begin{eqnarray}
W_{ \bf \bar{a} }(\vec{n}) \equiv
\sum_{ \vec{m} \in Z^3}
\delta_{\vec{n}, 2 \vec{m} + { \bf \bar{a} }} \  .
\end{eqnarray}
The cubic wall source operator for the pseudo-Goldstone
pions can then be expressed as follows
\begin{eqnarray}
{\cal O}_{\rm source} = \sum_{ \vec{n}, \vec{n}', { \bf \bar{a} } }
\bar{\chi}^a ( \vec{n}, t = 0) W_{ \bf \bar{a} }( \vec{n} )
\ \epsilon( { \bf \bar{a} }) \delta_{a,b}\ 
W^{\dagger}_{ \bf \bar{a} }(\vec{n}')
\chi^b ( \vec{n}', t = 0) \ ,
\end{eqnarray}
where $ \epsilon(x) \equiv (-1)^{x_1 + x_2 + x_3 + x_4} $,
$ a $ and $ b $ are color indices and
$ \chi $ is a staggered fermion field.
As an example, let us choose the sink operator to be 
the bilinear operator with spin structure $ S $
and flavor structure $ F $:
\begin{eqnarray}
{\cal O}_{\rm sink}(y) = \bar{\chi}(y_A)
\overline{ (\gamma_S \otimes \xi_F) }_{AB} \chi(y_B) \ .
\end{eqnarray}
$ y_A = 2 y + A $,
where $ y \in Z^4 $ is the coarse
lattice coordinate and
$ A \in \{0,1\}^4 $ is the hypercubic coordinate.
Then the correlation function is
\begin{eqnarray}
& & - \langle {\cal O}_{\rm sink}(y) \ {\cal O}_{\rm source} \rangle
\nonumber \\
& & \hspace*{5mm} = 
\sum_{ \vec{n}, \vec{n}', { \bf  \bar{a} } }
\overline{ (\gamma_S \otimes \xi_F) }_{AB} G^{c,a}(y_B, \vec{n})
W_{\bf \bar{a}}( \vec{n} )
\epsilon({\bf \bar{a}}) \delta_{a,b} W^{\dagger}_{ {\bf \bar{a}} }
( \vec{n}' ) G^{b,c}(\vec{n}', y_A)
\nonumber \\
& & \hspace*{5mm} =
\sum_{ \vec{n}, \vec{n}', { \bf \bar{a} } }
\overline{ (\gamma_5 \gamma_S \otimes \xi_5 \xi_F) }_{AB}
G^{c,a}(y_B, \vec{n}) W_{\bf \bar{a}}( \vec{n} ) \delta_{a,b}
W^{\dagger}_{\bf \bar{a}}( \vec{n}' ) 
[G^{c,b}(y_A, \vec{n}')]^*
\nonumber \\
& & \hspace*{5mm} = 
\overline{ (\gamma_5 \gamma_S \otimes \xi_5 \xi_F) }_{AB}
\sum_{ c,b,{\bf \bar{a}} }
\psi_{\bf \bar{a}}^{cb} (y_B)
[\psi_{\bf \bar{a}}^{cb} (y_A) ]^*
\label{final-eq}
\end{eqnarray}
where $ G (x,y) \equiv \frac{1}{2} [ G_P(x,y) + G_A(x,y)] $
for the domain $ 0 \le x_t, y_t < 2 N_t $ and
\begin{eqnarray}
\psi_{\bf \bar{a}}^{cb} (y_A) \equiv
\sum_{\vec{n}}
G^{c,b}(y_A, \vec{n}) W_{\bf \bar{a}}( \vec{n} ) \ .
\end{eqnarray}
Here $ \psi_{\bf \bar{a}}^{cb} (y_A) $ is actually
what we calculate on the computer
using a conjugate gradient method with
the wall source set to $ W_{\bf \bar{a}} $. 
For the $ f_K $ and the vacuum saturation part of $ B_K $,
Eq. (\ref{final-eq}) is used in our numerical simulation.
The idea of the above example (bilinear sink operators)
can be extended as a whole to the four fermion operator measurements
(for example, $ B_K $)
without loss of generality.
\section{Operators Computed}
Here, we present a set of lattice operators
which describe the same physics as
the continuum $ \Delta S =2 $ operator
and which have been used for our numerical
calculation of $ B_K $.
In the continuum, $ B_K $ is defined as
\begin{eqnarray}
\label{bk-def}
B_K(\mu) \equiv \frac
{
\langle \bar{K}^0 \mid  \bar{s} \gamma_{\mu}
(1-\gamma_5) d \  \bar{s} \gamma_{\mu}
(1-\gamma_5) d \mid K^0 \rangle
}{
\frac{8}{3} \langle \bar{K}^0 \mid
\bar{s} \gamma_{\mu} \gamma_5 d \mid 0 \rangle
\langle 0 \mid \bar{s} \gamma_{\mu} \gamma_5 d
\mid K^0 \rangle \    .
}
\end{eqnarray}
The numerator of $ B_K $ is a matrix element
of the $ \Delta S = 2 $ four-fermion operator with
the neutral $ K $ meson (kaon) states.
The denominator of $ B_K $ represents the vacuum saturation
approximation of the numerator, which inserts the 
vacuum state between the two bilinears of the
$ \Delta S = 2 $ four-fermion operators.
For the denominator, one needs to prescribe
a lattice bilinear operator
which corresponds to the continuum axial current:
\begin{eqnarray}
A_{\mu} = \bar{\chi}_s ( \overline{
\gamma_{\mu 5} \otimes \xi_5 } ) \chi_d \ .
\end{eqnarray}
Here the notations are the same as those adopted by Ref.
\cite{sharpe0,sharpe1,wlee0}.
The axial current is chosen
such that the flavor structure is identical to that of the
pseudo Goldstone kaon of the exact $ U_A(1) $ symmetry
in the staggered fermion action. 
For the numerator, we need some particular set of
lattice four-fermion operators
which correspond to the continuum $ \Delta S = 2 $ operator. 
There are two methods to transcribe  
the continuum $ \Delta S = 2 $ operator
on the lattice  with staggered fermions:
the one spin trace formalism and the two spin trace formalism
\cite{sharpe0,sharpe1,wlee0}.
In the continuum, there is no difference between the one spin
trace form and the two spin trace form of the $ \Delta S = 2 $
operator, since they are connected by the Fierz  transformation.
However, in the staggered fermion method, the one spin trace form
is different from the two spin trace form due to a pure lattice
artifact of 4 degenerate flavors.
In the {\em two spin trace formalism},
the four--fermion operator in the numerator
in Eq. (\ref{bk-def}) is transcribed to the lattice
\cite{sharpe0,sharpe1,wlee0}
as a sum of four operators:
\begin{eqnarray}
{\cal O}^{Latt}_{2TR} & = &
(V \times P)^{2TR}_{ab;ba}
+ (V \times P)^{2TR}_{aa;bb}
\nonumber \\
& & \label{two-tr-op}
+ (A \times P)^{2TR}_{ab;ba}
+ (A \times P)^{2TR}_{aa;bb}
\end{eqnarray}
where V (or A) represents the vector (or axial) spin structure,
P represents the pseudoscalar-like flavor structure and
the subscript $ ab;ba $ (or $aa;bb$) represents
the color indices of the quark fields
(the details of these notations are described 
in Ref. \cite{wlee0}).
The operators in Eq. (\ref{two-tr-op}) have the same
chiral behavior in the limit of vanishing quark mass
as the continuum $ \Delta S = 2 $ operator
\cite{sharpe1,wlee0,sharpe30}.
In addition, the matrix elements of $ {\cal O}^{Latt}_{2TR} $
show the same leading logarithmic dependence
on the renormalization scale as
the continuum $ \Delta S = 2 $ operator
\cite{wlee0,japan1,sharpe4}.
In the {\em one spin trace formalism},
the four-fermion operator of the numerator
in Eq. (\ref{bk-def}) is transcribed to the
lattice as follows:
\begin{eqnarray}
{\cal O}^{Latt}_{1TR} & = &
(V \times P)^{1TR}_{ab;ba}
+ (V \times P)^{1TR}_{aa;bb}
\nonumber \\
& &
+ (A \times P)^{1TR}_{ab;ba}
+ (A \times P)^{1TR}_{aa;bb}
\nonumber \\
& &\label{one-tr-op}
+ {\cal O}^{1TR}_{\rm chiral \  partner} \ .
\end{eqnarray}
where again the details of this notation is given in Ref. \cite{wlee0}.
In contrast with the {\em two spin trace formalism},
the individual terms in Eq. (\ref{one-tr-op})
do not possess the same chiral behavior
as the continuum $ \Delta S = 2 $ operator \cite{wlee0}.
We must add $ {\cal O}^{1TR}_{\rm chiral \   partner} $
in order to preserve the correct continuum chiral behavior
\cite{wlee0}.
By imposing the correct chiral behavior on
$ {\cal O}^{Latt}_{1TR} $,
we determine the chiral partner
operator \cite{wlee0} as follows:
\begin{eqnarray}
{\cal O}^{1TR}_{\rm chiral \  partner} & = &
(V \times S)^{1TR}_{ab;ba}
+ (V \times S)^{1TR}_{aa;bb}
\nonumber \\
& &
\hspace*{-0.25 in}
+ (A \times S)^{1TR}_{ab;ba}
+ (A \times S)^{1TR}_{aa;bb}\ .
\label{chi-partner}
\end{eqnarray}
This forces the resulting operator to respect the continuum
chiral behavior.
The next question is whether
the whole operators including the additional chiral partner operators 
still have the continuum leading logarithmic behavior.
First of all, we need to choose
the basis operators such that they belong to
the identity representation with respect to the
90$^{\circ}$ axial rotation group (a subgroup
of the exact $ U_A(1) $ symmetry group).
This particular choice of the basis operators guarantees the analytic
chiral behavior of the continuum.
Second, we find that  an eigen-operator
(Eq. (\ref{one-tr-op}) and
Eq. (\ref{chi-partner})) possesses
the same leading logarithmic behavior
as the continuum $ \Delta S = 2 $ operator
\cite{wlee0}.
For $ B_K $ measurements, 
the difference between the one spin trace operators
and the two spin trace operators vanishes as
$ a \rightarrow 0 $ \cite{wlee0}.
We have used both one spin trace and two spin trace operators
to calculate $ B_K $ on the lattice 
and  the numerical results are compared later
in this paper.
\section{Simulations and Measurement Parameters}
In this section we describe
the simulation parameters for the gauge configurations
and $ B_K $ measurements. 
Our old hadron mass calculation \cite{brown}
on a $ 16^3 \times 32 $ lattice
at $ \beta = 5.7 $ informed us
that a longer time dimension allows a more precise
fitting of the hadron propagators' exponential
time dependence.
For this reason the volume of the configuration was chosen as
$ 16^3 \times 40 $. The coupling constant was
$ \beta $ = 5.7~ ($ 1/a \approx 2 $ GeV).
The dynamical effects of two degenerate
flavors of staggered fermions with a mass 0.01
were incorporated into the gauge configurations,
using the hybrid molecular dynamics R-algorithm
\cite{gottlieb}.
The sea quark mass (0.01) has been fixed
through all the measurements even though
various valence quark masses
were chosen for the $ B_K $ measurement.
The gauge configurations were updated
by the hybrid molecular dynamics R-algorithm
with molecular dynamics
step size 0.0078125 \cite{gottlieb} and a trajectory
length of 0.5 time units.
Now the measurement parameters for
$ B_K $:
Every 60 trajectories
$ B_K $ has been measured.
The total number of the gauge configuration samples for
$ B_K $ measurements was 155.
We have used both cubic wall source and 
conventional even-odd wall source methods to
create the pseudo Goldstone boson.
We used two separate wall sources 
to create $ K^0 $ and $ \bar{K}^0 $ mesons.
The distance between these two separate wall sources
was 36 lattice units.
For each $ B_K $ measurement,
both wall sources were shifted by 15 lattice units
in the time direction from the position used in the previous
measurement,
while the distance between them
was fixed to 36 through all the measurements.
The valence (quark, antiquark) mass pairs
for the $ K $ mesons were (0.01, 0.01),
(0.02, 0.02), (0.03, 0.03), (0.004, 0.01), (0.004, 0.02),
(0.01, 0.03) and (0.004, 0.05).
The quark propagators were calculated,
using the lattice doubling method
described in the previous section.
The stopping condition of the conjugate gradient
residual for the quark propagator 
was set to $ 1.0 \times 10^{-8} $.
The quark propagators were gauge fixed to
Landau condition with a numerical precision of
$ a^4 g^2 ( \partial_{\mu} A_{\mu} )^2  < 1.0 \times 10^{-7} $.
\section{Data Analysis}
We now present the numerical
results of $ {\cal M}_K $ (the numerator
of $ B_K $) and the vacuum saturation amplitude
$ {\cal M}^V_K $ (the denominator
of $ B_K $) with respect to the lattice
Euclidean time for Kaon
with a quark anti-quark mass pair = (0.01, 0.01).
Meanwhile, we explain how the
central plateau region has been selected
to determine $ B_K $. 
We  also discuss the numerical results for 
unrenormalized (naive, bare) $ B_K $
with respect to the various quark and anti-quark masses.
In addition, we will present the numerical
measurement of the wrong flavor channel
$ ((V+A) \otimes S)^{2TR} $ in order to see
how much contamination comes from the operator
mixing and the excited hadronic states
which are supposed to vanish in the limit
of $ a \rightarrow 0 $.
\subsection{The Denominator: Vacuum Saturation Amplitude}
\begin{figure}[t]

\centering

\epsfxsize=\hsize\epsfbox{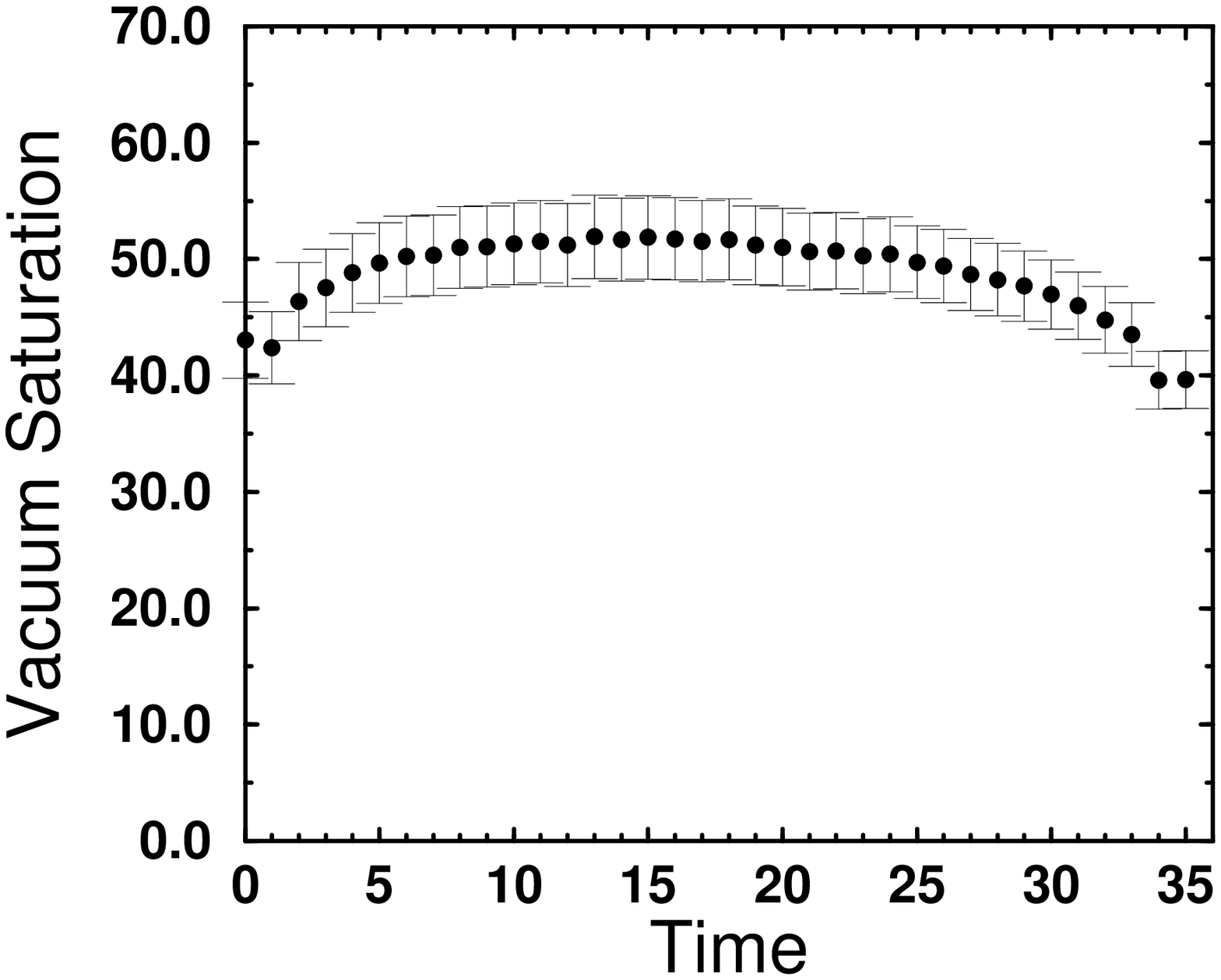}

\caption{Vacuum saturation with respect to time.
$ m_{\rm d} a = m_{\rm s} a $ = 0.01.
Calculated with the even-odd source
method in the two spin trace formalism.
}

\label{fig:1}

\end{figure}

\begin{figure}[t]
 
\centering

\epsfxsize=\hsize\epsfbox{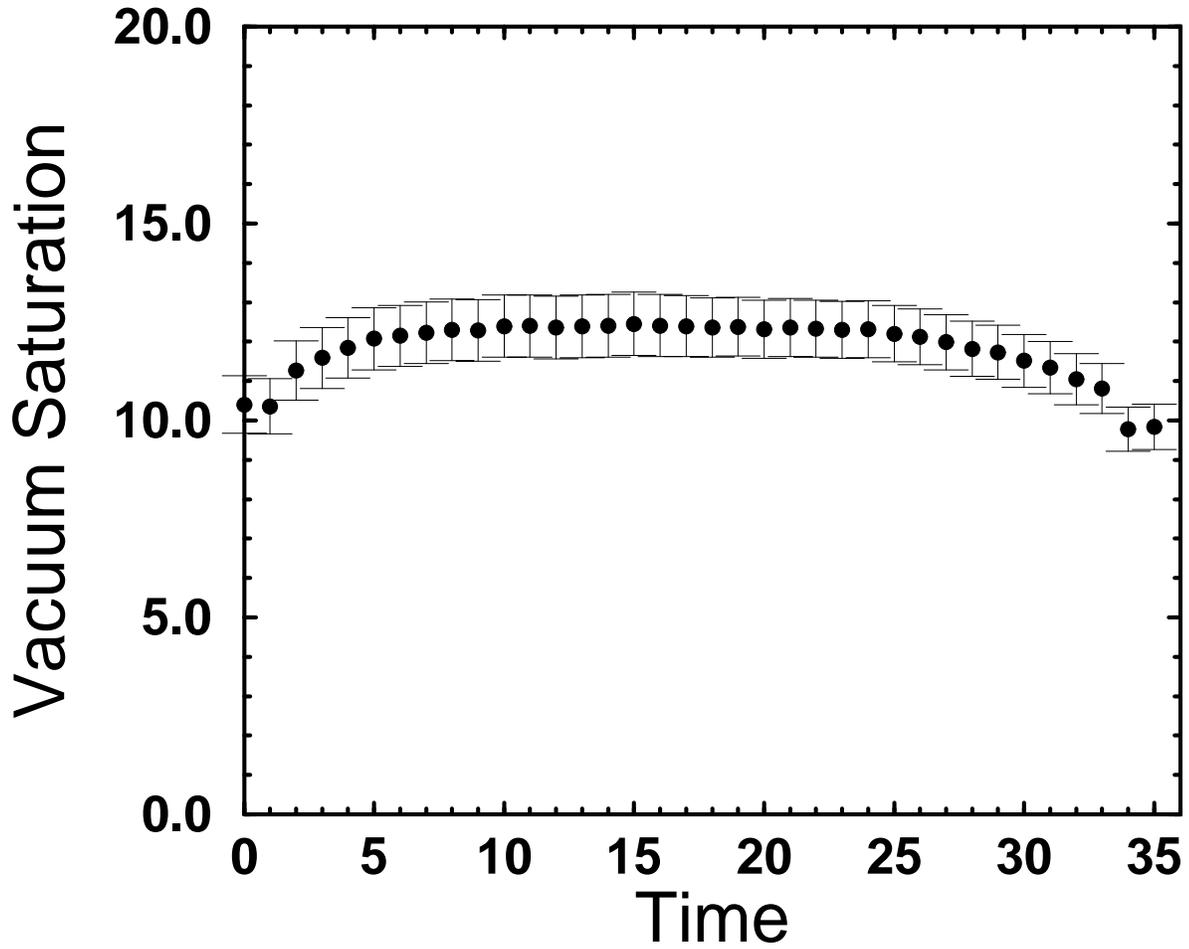}

\caption{Vacuum saturation with respect to time.
$ m_{\rm d} a = m_{\rm s} a $ = 0.01.
Calculated with the cubic wall source
method in the two spin trace formalism.
}

\label{fig:2}
 
\end{figure}

We define
the denominator of $ B_K $
as the vacuum saturation amplitude:
\begin{equation}
{\cal M}^V_K \equiv
\frac{8}{3} \langle \bar{K}^0 \mid
\bar{s} \gamma_{\mu} \gamma_5 d \mid 0 \rangle
\langle 0 \mid \bar{s} \gamma_{\mu} \gamma_5 d
\mid K^0 \rangle \ .
\end{equation}
Numerical data for $ {\cal M}^V_K $ 
is presented in Figure \ref{fig:1} (the even-odd
wall source method) and in Figure \ref{fig:2}
(the cubic wall source method).
The data points in Figure \ref{fig:1}, \ref{fig:2}, and
\ref{fig:3} are obtained by single elimination jack-knife method.
Each data point in  Figure \ref{fig:1}(the even-odd wall source)
has about twice larger error than that in
Figure \ref{fig:2} (the cubic wall source).
However, one needs to notice that 
the cubic wall source method takes four times
longer time to compute than the even-odd wall source method.
\begin{table}[t]
\renewcommand{\baselinestretch}{1}
\begin{center}
\small
\begin{tabular}{||c||c|c|c|c||} \hline \hline
 & \multicolumn{2}{c|}{Cubic Wall Source} 
 & \multicolumn{2}{c||}{Even-Odd Wall Source} 
\\ \cline{2-5}
Fitting Range & 
$ {\cal M}^V_K $ & 
$ \chi^2/(dof) $ & 
$ {\cal M}^V_K $ & 
$ \chi^2/(dof) $ 
\\ \hline \hline
$ 14 \le t \le 21 $ &
12.17(66) & 0.40(63) & 49.2(30) & 0.40(68)
\\ \hline
$ 13 \le t \le 22 $ &
12.10(58) & 0.37(52) & 49.3(25) & 0.33(53)
\\ \hline
$ 12 \le t \le 23 $ &
12.15(60) & 0.37(48) & 48.7(24) & 0.61(65)
\\ \hline
$ 11 \le t \le 24 $ &
12.10(63) & 0.45(48) & 49.0(25) & 0.68(56)
\\ \hline
$ 10 \le t \le 25 $ &
11.76(59) & 1.06(51) & 47.0(27) & 1.24(76)
\\ \hline
$ 9 \le t \le 26 $ &
11.70(59) & 1.27(66) & 46.7(27) & 1.34(70)
\\ \hline
$ 8 \le t \le 27 $ &
11.61(58) & 1.31(62) & 45.2(24) & 1.40(68)
\\ \hline
$ 7 \le t \le 28 $ & 
11.20(58) & 1.65(70) & 44.2(24) & 1.52(70)
\\ \hline
$ 6 \le t \le 28 $ &
10.74(60) & 1.72(65) & 42.9(24) & 1.47(67)
\\ \hline \hline
\end{tabular}
\caption{Here, we present the numerical results
for  the vacuum saturation amplitude ($ {\cal M}^V_K $) with
the quark mass pair (0.01, 0.01), calculated
both in the even-odd wall source method and in
the cubic wall source method.
All the values in the table have been obtained through
the covariance fitting to a constant
on the bootstrap ensembles.}
\label{vs-table}
\end{center}
\end{table}
%
%
In Table \ref{vs-table},
the results of the
covariance fitting of
the vacuum saturation
together with their $ \chi^2 $ per degree of freedom
are given with respect to the various fitting ranges
in the lattice Euclidean time for quark mass $ m_q a = 0.01 $.
From Table \ref{vs-table},
for the cubic wall source method
the minimum of the $ \chi^2 $ per degree of freedom
extends to the fitting range of $ 11 \le t \le 24 $.
For the even-odd wall source method, 
the minimum of the $ \chi^2 $ per degree of freedom
occurs in the fitting range of $ 13 \le t \le 23 $.
It is important to keep in mind the fact that 
both the one spin trace form and the two spin trace form
of $ B_K $ have the vacuum saturation in common.
\subsection{The Numerator: $ {\cal M}_K $}
\begin{figure}[t]

\centering

\epsfxsize=\hsize\epsfbox{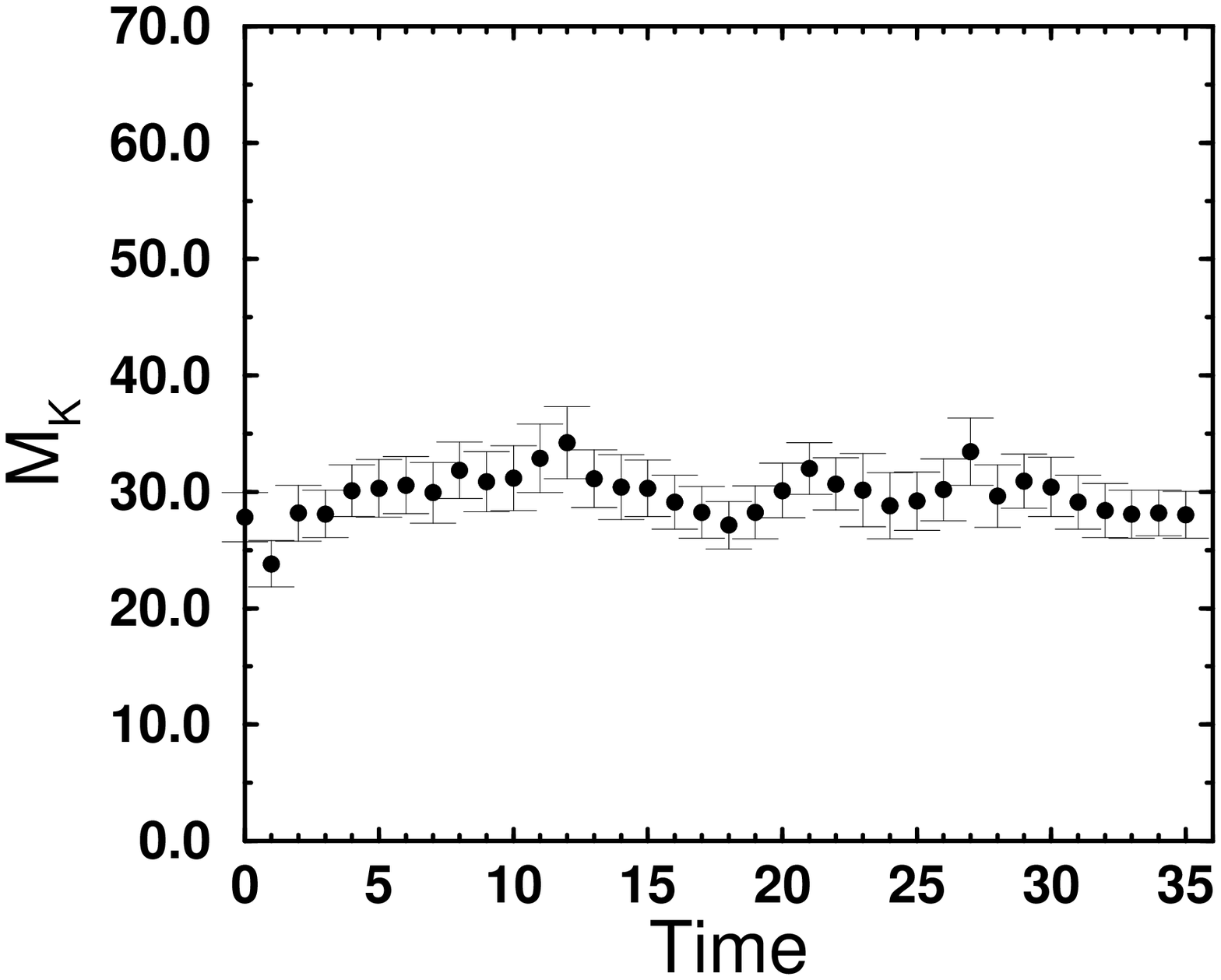}

\caption{$ M_K $ with respect to  time.
$ m_{\rm d} a = m_{\rm s} a $ = 0.01.
Calculated with the even-odd wall source
method in the two spin trace formalism.}

\label{fig:3}

\end{figure}

\begin{figure}[t]
 
\centering

\epsfxsize=\hsize\epsfbox{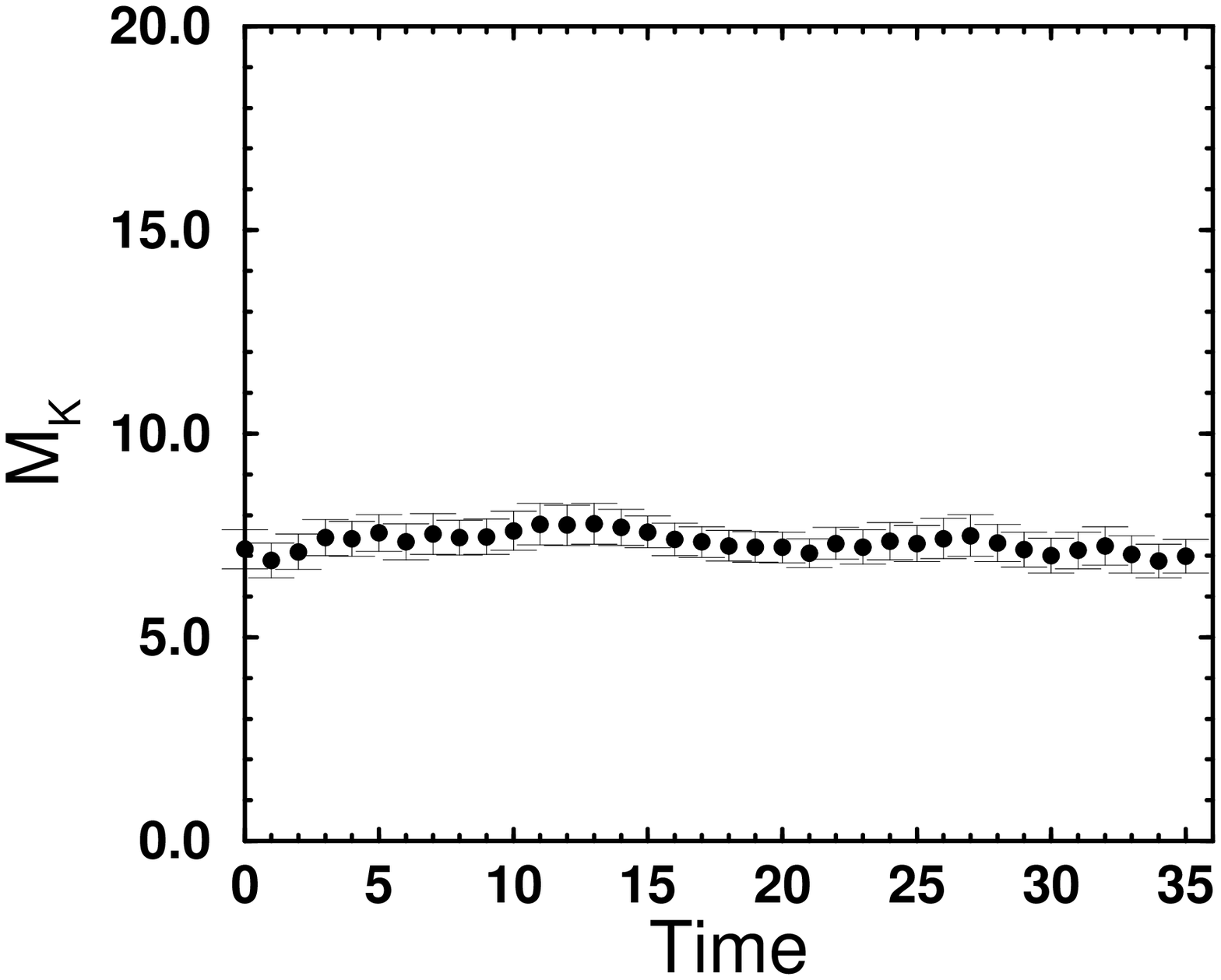}
 
\caption{$ M_K $ with respect to time.
$ m_{\rm d} a = m_{\rm s} a $ = 0.01.
Calculated with the cubic wall source
method in the two spin trace formalism.}
 
\label{fig:4}

\end{figure}

\begin{figure}[t]
 
\centering

\epsfxsize=\hsize\epsfbox{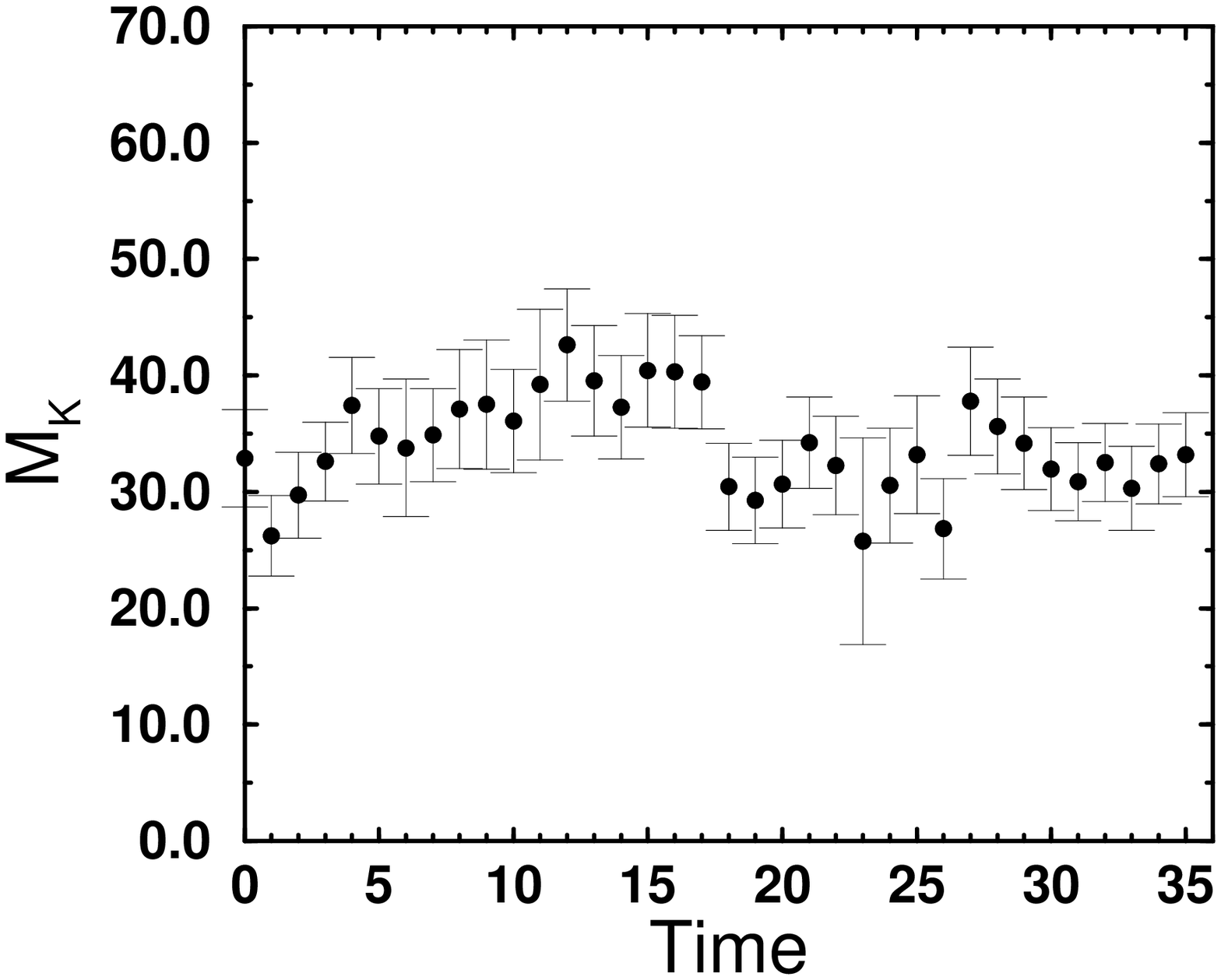}
 
\caption{$ M_K $ with respect to time.
$ m_{\rm d} a = m_{\rm s} a $ = 0.01.
Calculated with the even-odd wall source
method in the one spin trace formalism.}
 
\label{fig:5}
 
\end{figure}

We define the numerator of $ B_K $ 
as:
\begin{equation}
{\cal M}_K \equiv \langle \bar{K}^0 \mid  \bar{s} \gamma_{\mu}
(1-\gamma_5) d \  \bar{s} \gamma_{\mu}
(1-\gamma_5) d \mid K^0 \rangle
\end{equation}
Numerical data for $ {\cal M}_K $ calculated in 
the two spin trace formalism with the even-odd wall
source method is presented in Figure \ref{fig:3}.
Numerical data for $ {\cal M}_K $ calculated in
the two spin trace formalism with the cubic wall source
method is shown in Figure \ref{fig:4}.
Numerical data for $ {\cal M}_K $ calculated in
the one spin trace formalism  with the even-odd wall
source method is drawn in Figure \ref{fig:5}.
All the data points in Figures \ref{fig:3}, \ref{fig:4}, and \ref{fig:5}
are obtained by the single elimination jack-knife method.
Each data point in  Figure \ref{fig:3} (the even-odd wall source)
has an about twice larger error than that in
Figure \ref{fig:4} (the cubic wall source).
The nearby data points
in Figure \ref{fig:3} and Figure \ref{fig:5} (the even-odd wall source)
have larger fluctuations than those in Figure \ref{fig:4}
(the cubic wall source), which are reflected on 
$ \chi^2 / \rm d.o.f.$  in Table \ref{mk-table}. 
However, this was not observed in
the results for vacuum saturation amplitude.
It is important to note
the fact that the computational time for the
cubic wall source method is four times longer than
that for the even-odd wall source method.
\begin{table}[t]
\renewcommand{\baselinestretch}{1}
\begin{center}
%
\footnotesize
\begin{tabular}{||c||c|c|c|c|c|c||} \hline \hline
 & \multicolumn{4}{c|}{Two Spin Trace Form}
 & \multicolumn{2}{c||}{One Spin Trace Form}
\\ \cline{2-7}
Fitting & \multicolumn{2}{c|}{Cubic Wall Source}
 & \multicolumn{2}{c|}{Even-odd Wall Source} 
 & \multicolumn{2}{c||}{Even-odd Wall Source}
 \\ \cline{2-7}
Range & 
$ {\cal M}_K $ & 
$ \chi^2/(\rm d.o.f.) $ & 
$ {\cal M}_K $ & 
$ \chi^2/(\rm d.o.f.) $ & 
$ {\cal M}_K $ & 
$ \chi^2/(\rm d.o.f.) $ 
\\ \hline \hline
$ 14 \le t \le 21 $ &
7.20(48) & 1.11(99) & 29.8(15) & 1.05(102) & 33.4(21) & 1.45(106)
\\ \hline
$ 13 \le t \le 22 $ &
7.23(29) & 0.88(87) & 29.9(15) & 0.87(83) & 33.6(17) & 1.18(82)
\\ \hline
$ 12 \le t \le 23 $ &
7.26(28) & 0.75(75) & 28.7(14) & 1.47(80) & 34.0(17) & 1.19(68)
\\ \hline
$ 11 \le t \le 24 $ &
7.19(28) & 0.73(67) & 28.6(13) & 1.24(70) & 34.1(18) & 1.03(62)
\\ \hline
$ 10 \le t \le 25 $ &
7.16(29) & 0.66(60) & 28.1(14) & 1.29(59) & 33.7(16) & 0.96(58)
\\ \hline
$ 9 \le t \le 26 $ &
7.11(30) & 0.62(51) & 28.1(13) & 1.14(54) & 33.7(16) & 1.04(49)
\\ \hline
$ 8 \le t \le 27 $ &
7.11(30) & 0.60(52) & 27.7(12) & 1.27(61) & 33.6(16) & 0.95(47)
\\ \hline
$ 7 \le t \le 28 $ & 
7.04(27) & 0.57(54) & 27.1(11) & 1.47(52) & 33.9(16) & 0.95(43)
\\ \hline
$ 6 \le t \le 29 $ &
6.97(26) & 0.61(50) & 27.3(12) & 1.52(54) & 33.8(15) & 0.90(45)
\\ \hline \hline
\end{tabular}
\caption{Here, we present the numerical results
for the $ {\cal M}_K $ with
the quark mass pair (0.01, 0.01), calculated
both in the even-odd wall source method and in
the cubic wall source method.
All the values in the table has obtained through
the covariance matrix fitting to a constant 
on the bootstrap ensembles.}
\label{mk-table}
\end{center}
\end{table}
In Table \ref{mk-table},  the covariance matrix
fitting results of $ {\cal M}_K $
and its $ \chi^2/(\rm d.o.f.) $ are collected
with respect to the fitting range in
the lattice Euclidean time.
From Table \ref{mk-table}
it is difficult to choose
the optimal fitting range
as all of the $ \chi^2/(\rm d.o.f.) $
values are within $ 0.5 \sim 1 $ $ \sigma $ for
$ {\cal M}_K $ calculated
in the two spin trace form with the cubic wall source.
This is also true for $ {\cal M}_K $ calculated
in the one spin trace form with the even-odd wall source.
For these cases, we therefore choose the optimal fitting range consistent 
with the vacuum saturation amplitude $ {\cal M}^V_K $.

For $ {\cal M}_K $ calculated in the two spin trace form 
with the even-odd wall source, we notice that
the optimal fitting range is $ 13 \le t \le 22 $.
The $ {\cal M}_K $ results in the cubic wall source method
has more correlation (less fluctuation) between neighboring 
time slices than those in the even-odd wall source method,
while this is not obvious for the measurements 
of vacuum saturation amplitude $ {\cal M}^V_K $.
\subsection{The Ratio: $ B_K $}
\begin{table}[p]
\renewcommand{\baselinestretch}{1}
\begin{center}
%
\footnotesize
\begin{tabular}{||c||c|c||c||} \hline \hline
 & \multicolumn{3}{c||}{Fitting Range} 
\\ \cline{2-4}
Quark Mass & \multicolumn{2}{c||}{Two Spin Trace Form}
 & One Spin Trace Form
\\ \cline{2-4}
Pair & Cubic Wall Source
 & Even-odd Wall Source 
 & Even-odd Wall Source
\\ \hline \hline
(0.004, 0.01) &
$ 9 \le t \le 26 $  &  $ 13 \le t \le 22 $  &  $ 13 \le t \le 22 $
\\ \hline
(0.004, 0.02) &
$ 11 \le t \le 24 $  &  $ 13 \le t \le 22 $  &  $ 13 \le t \le 22 $
\\ \hline
(0.004, 0.05) &
$ 14 \le t \le 21 $  &  $ 15 \le t \le 20 $  &  $ 15 \le t \le 20 $ 
\\ \hline
(0.01, 0.01) &
$ 11 \le t \le 24 $  &  $ 13 \le t \le 22 $  &  $ 13 \le t \le 22 $
\\ \hline
(0.01, 0.03) & 
$ 11 \le t \le 24 $  &  $ 13 \le t \le 22 $  &  $ 13 \le t \le 22 $
\\ \hline
(0.02, 0.02) &
$ 13 \le t \le 22 $  &  $ 13 \le t \le 22 $  &  $ 13 \le t \le 22 $
\\ \hline
(0.03, 0.03) &
$ 13 \le t \le 22 $  &  $ 13 \le t \le 22 $  &	$ 13 \le t \le 22 $
\\ \hline \hline
\end{tabular}
\caption{Here, we present the optimal fitting
range with respect to quark mass pairs.
The optimal fitting range implies that
we can get the same average with smaller error bar
for $ \chi^2 \le 1.0 $.
}
\label{fit-range-table}
\end{center}
\end{table}

\begin{figure}[t]

\centering

\epsfxsize=\hsize\epsfbox{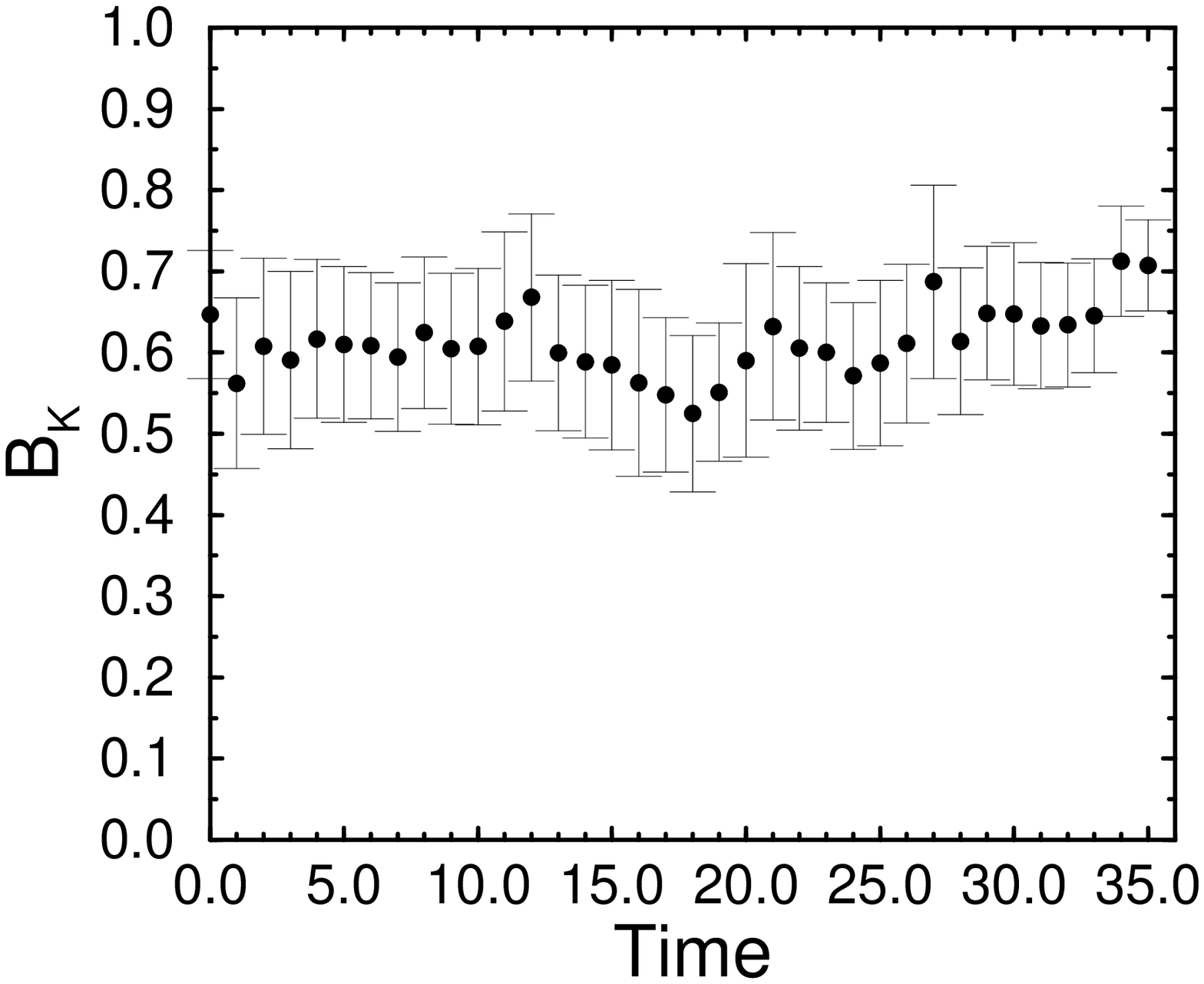}

\caption{Unrenormalized $ B_K $ with respect to time.
$ m_{\rm d} a = m_{\rm s} a $ = 0.01.
Calculated with the even-odd wall
source method in the two spin trace formalism.}

\label{fig:6}

\end{figure}

\begin{figure}[t]
 
\centering

\epsfxsize=\hsize\epsfbox{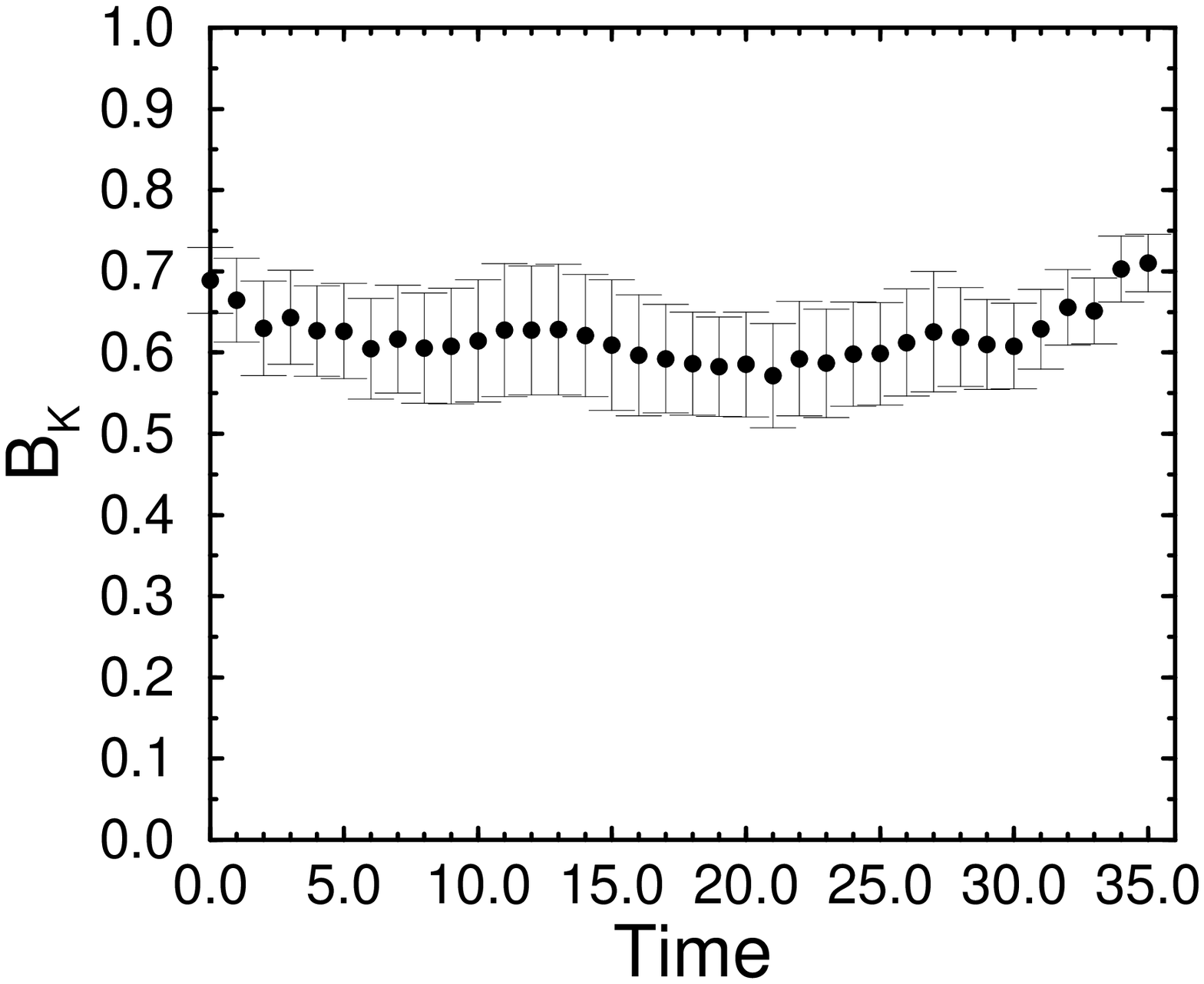}
 
\caption{Unrenormalized $ B_K $ with respect to time.
$ m_{\rm d} a = m_{\rm s} a $ = 0.01.
Calculated with the cubic wall
source method in the two spin trace formalism.}
 
\label{fig:7}

\end{figure}

\begin{figure}[t]

\centering

\epsfxsize=\hsize\epsfbox{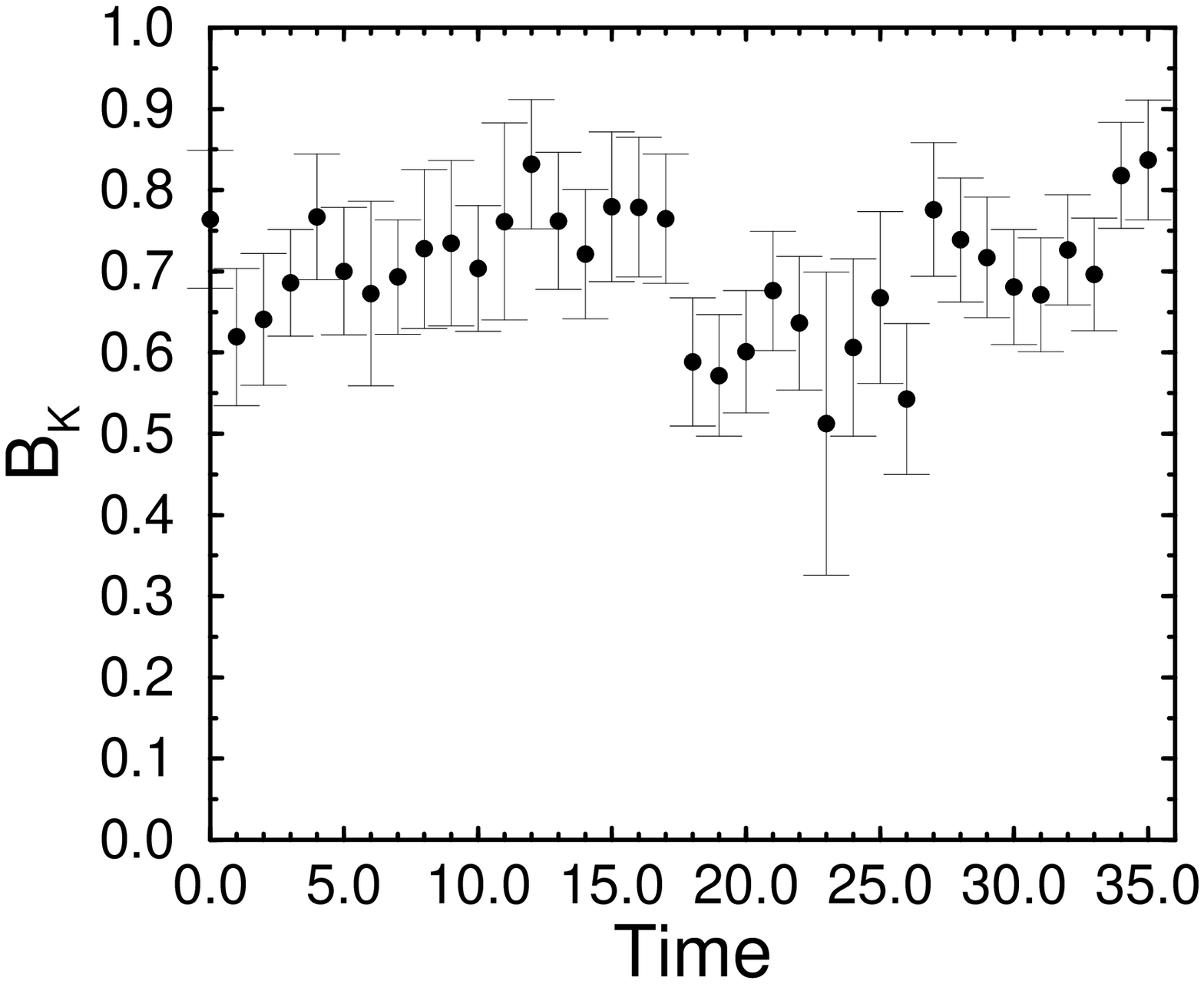}

\caption{Unrenormalized $ B_K $ with respect to time.
$ m_{\rm d} a = m_{\rm s} a $ = 0.01.
Calculated with the even-odd wall
source method in the one spin trace formalism.}

\label{fig:8}

\end{figure}

In the previous sections, we have discussed
the covariance matrix fitting result of $ {\cal M}_K $
and $ {\cal M}^V_K $ as a function of the fitting range.
From this analysis of the fitting ranges,
we determine the
optimal fitting range for $ B_K $.
The optimal fitting ranges we have chosen
to determine $ B_K $ with respect to various 
quark mass pairs are summarized 
in Table \ref{fit-range-table}.
Once the fitting range is chosen,
$ B_K $ can be determined through the covariance
fitting to a constant.
One of our fitting procedures is naive jack-knife
analysis of the data (conventional) and 
the other uses the covariance fitting
on the bootstrap ensembles.
The numerical results for $ B_K $ calculated
in the two spin trace form
with the even-odd wall source method
are shown in Figure \ref{fig:6}.
The numerical results for $ B_K $ calculated
in the two spin trace form
with the cubic wall source method
are drawn in Figure \ref{fig:7}.
The numerical results for $ B_K $ calculated
in the one spin trace form
with the even-odd wall source method
appear in Figure \ref{fig:8}.
\begin{table}[p]
\renewcommand{\baselinestretch}{1}
\begin{center}
%
\footnotesize
\begin{tabular}{||c|c||c|c||c||}
\hline \hline
& &  \multicolumn{3}{c||}{unrenormalized $ B_K $} 
\\ \cline{3-5}
Quark Mass & $ m_K $
& \multicolumn{2}{c||}{Two Spin Trace Form} &
One Spin Trace Form \\ \cline{3-5}
Pair &  & Cubic Source  & Even-odd Source &
Even-odd Source \\ \hline \hline
(0.004, 0.01) & 0.219(2) & 0.557(32) & 0.547(54) & 0.656(73)
\\ \hline
(0.004, 0.02) & 0.277(2) & 0.641(32) & 0.630(39) & 0.698(58)
\\ \hline
(0.004, 0.05) & 0.406(2) & 0.731(34) & 0.687(44) & 0.759(88)
\\ \hline
(0.01, 0.01) & 0.253(2) & 0.600(27) & 0.579(33) & 0.688(46)
\\ \hline
(0.01, 0.03) & 0.348(2) & 0.710(26) & 0.689(30) & 0.748(37)
\\ \hline
(0.02, 0.02) & 0.347(1) & 0.709(25) & 0.689(28) & 0.746(33)
\\ \hline
(0.03, 0.03) & 0.421(1) & 0.768(23) & 0.753(26) & 0.781(28)
\\ \hline \hline
\end{tabular}
\caption{$ K $ meson mass and unrenormalized $ B_K $ 
versus quark mass pair: the $ B_K $ data are analyzed 
by the single-elimination jack-knife method
over the optimal fitting range.
The $ K $ meson mass, $ m_K $ was obtained by analyzing the
results for $ \langle 0 \mid \bar{S} (\gamma_0 \gamma_5
\otimes \xi_5) D \mid K \rangle $.
}
\label{bk-table}
\end{center}
\end{table}
\begin{table}[p]
\renewcommand{\baselinestretch}{1}
\begin{center}
%
\footnotesize
\begin{tabular}{||c||c|c||c||}
\hline \hline
&  \multicolumn{3}{c||}{Unrenormalized $ B_K $} 
\\ \cline{2-4}
Quark Mass 
& \multicolumn{2}{c||}{Two Spin Trace Form} &
One Spin Trace Form \\ \cline{2-4}
Pair &  Cubic Source  & Even-odd Source &
Even-odd Source \\ \hline \hline
(0.004, 0.01) & 0.546(23) & 0.548(42) & 0.675(63)
\\ \hline
(0.004, 0.02) & 0.627(28) & 0.663(37) & 0.700(47)
\\ \hline
(0.004, 0.05)  & 0.732(30) & 0.715(40) & 0.771(78) 
\\ \hline
(0.01, 0.01)  & 0.595(22) & 0.607(34) & 0.682(45)
\\ \hline
(0.01, 0.03)  & 0.717(23) & 0.708(31) & 0.771(37)
\\ \hline
(0.02, 0.02) & 0.727(24) & 0.707(29) & 0.773(36)
\\ \hline
(0.03, 0.03) & 0.791(21) & 0.758(26) & 0.804(25)
\\ \hline \hline
\end{tabular}
\caption{
Unrenormalized $ B_K $ for each quark mass pair.
The $ B_K $ data is analyzed by the covariance
fitting over the optimal fitting range on the bootstrap
ensembles. 
}
\label{bk-cov-table}
\end{center}
\end{table}

Table \ref{bk-table} is a collection of 
the numerical results of $ K $ meson mass and 
$ B_K $ with respect to the
quark mass pairs calculated in the optimal fitting range
by the jack-knife method.
The results for $ K $ meson mass ($m_K$)
are obtained by analyzing
the results for axial current matrix element with an external
$ K $ meson state 
$ \langle 0 \mid \bar{S} (\gamma_0 \gamma_5 \otimes \xi_5) D
\mid K \rangle $,
which are also used to obtain the vacuum
saturation amplitude.
We fitted the logarithm of the axial current data to the
linear function of the Euclidean time $ A + m_K t $
in the optimal fitting range.
We would like to thank Pavlos Vranas for checking these mass results
using his own fitting program.
Table \ref{bk-cov-table} presents
the results of  $ B_K $  
obtained by the covariance fitting over the optimal fitting range
on the bootstrap ensembles.
\subsection{The Wrong Flavor Channel}
\begin{figure}[t]

\centering

\epsfxsize=\hsize\epsfbox{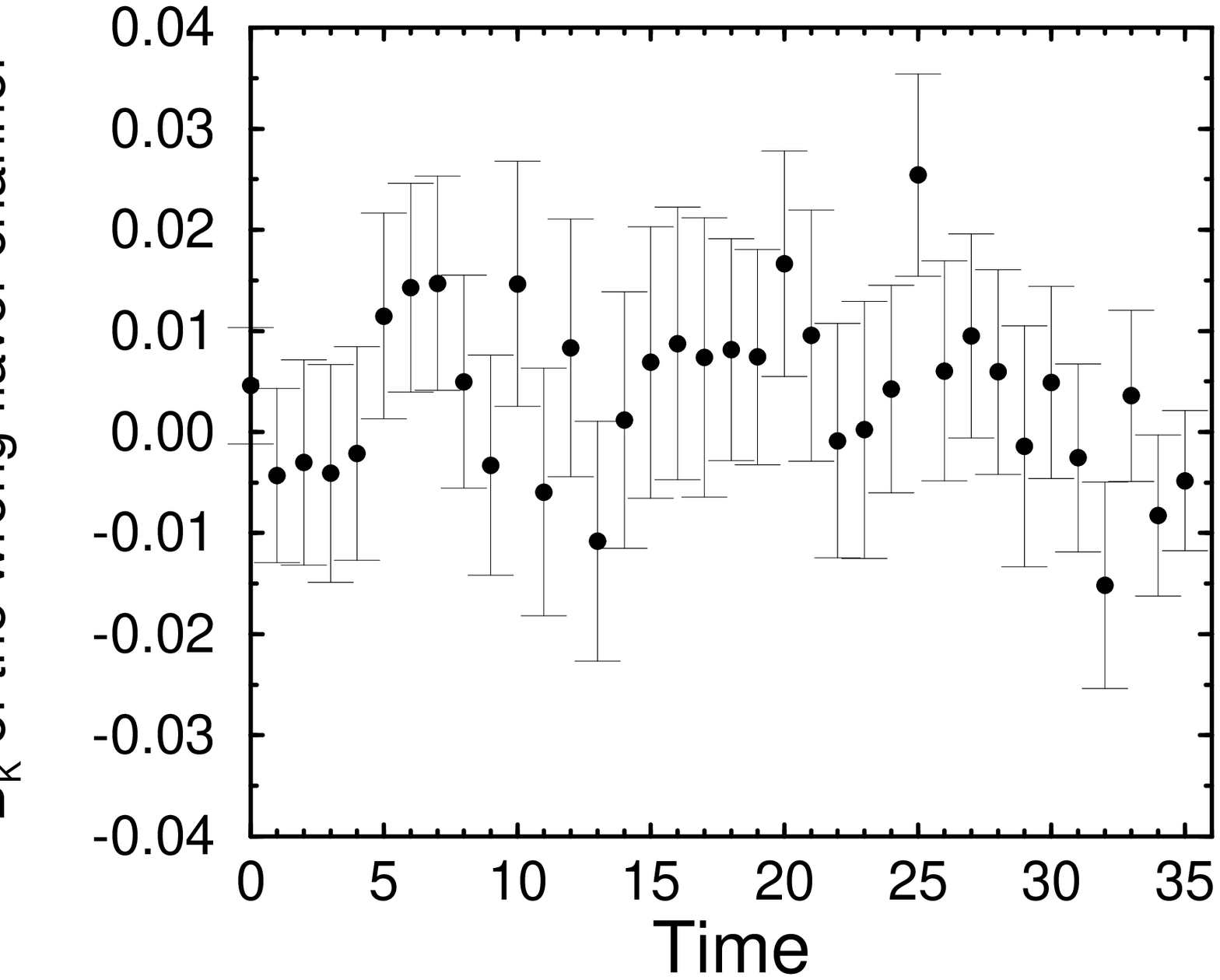}

\caption{Unrenormalized $ B_K $ with the wrong flavor
structure ($ ((V + A) \otimes S)^{\rm 2TR} $
with respect to time. $ m_d a = m_s a = 0.02 $.
Calculated with the even-odd
wall source method in the two spin trace
formalism.} 

\label{fig:9}

\end{figure}

\begin{figure}[t]

\centering

\epsfxsize=\hsize\epsfbox{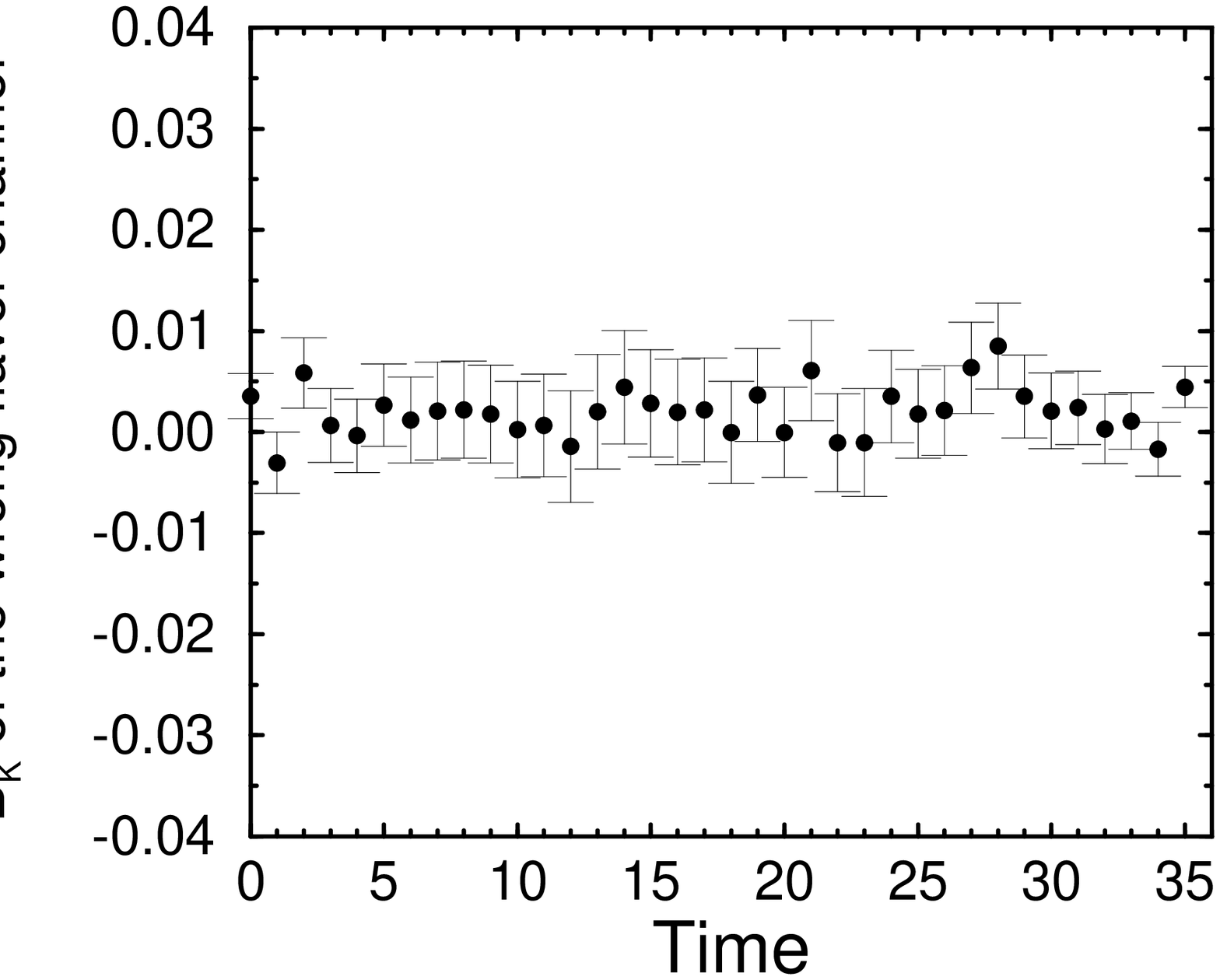}

\caption{Unrenormalized $ B_K $ with the wrong flavor
structure ($ ((V + A) \otimes S)^{\rm 2TR} $
with respect to time. $ m_d a = m_s a = 0.02 $.
Calculated with the cubic
wall source method in the two spin trace
formalism.} 

\label{fig:10}

\end{figure}

Here we address two important questions
on the validity of our approach to $ B_K $.
The first question comes from the fact that
the higher-loop radiative correction of the four-fermion
composite operators cause the violation of the
continuum spin and flavor symmetries.
It is important to know non-perturbatively,
how large is the contribution of such symmetry violating terms
to the weak matrix element measurements 
for finite lattice spacing. 
Note, such terms are supposed to vanish in the
limit of $ a \rightarrow 0 $.
The smaller the contribution from
these wrong flavor channels is, 
the more reliable our connection between the continuum
and lattice observables (either at tree level
or at one loop level).
The second question is how exclusively we can select
the pseudo-Goldstone mode through our improved wall source
method.
In other words, we want to ask how reliably our wall source
technique suppresses the unwanted hadronic states.
We have chosen the $ ((V+A) \otimes S)^{\rm 2TR} $ operator
in order to address the above two questions.
The matrix element of this operator with external $ K $ 
mesons is supposed to vanish in the continuum limit 
($ a \rightarrow 0 $) of lattice QCD, 
due to the vanishing flavor trace.
The numerical results for this wrong flavor channel
are shown in Figure \ref{fig:9} for the even-odd wall source
and Figure \ref{fig:10} for the cubic wall source.
From Figures \ref{fig:9} and \ref{fig:10}, we notice that the value of
the wrong flavor channel is extremely
suppressed (less than 1\% of $ B_K $) in both cases.
This implies that the unwanted operator
mixing of $ ((V+A) \otimes S)^{\rm 2TR} $ 
should be at most 1\% of $ B_K $
since it is suppressed by $ \alpha_s/(4\pi) $
as well as by the vanishing flavor trace.
This fact that the unwanted operator
mixing is smaller than 1\% of $ B_K $
is of great significance to our approximate matching
between the continuum and lattice composite operators.
So far, one has neglected those terms of wrong flavor channels
which enters at order $ g^2 $, when one connects the lattice
$ B_K $ to the continuum $ B_K $ at one loop level
\cite{sharpe1,wlee0,japan0,japan1}.
The main reason was that these wrong flavor channels
will not contribute  to $ B_K $
at all as $ a \rightarrow 0 $.
Hence, the remaining difficulty was to know 
how large is the contribution
of those wrong flavor channels at finite lattice spacing.
Our non-perturbative measurements of one of wrong flavor
channels shows that the contribution from these wrong flavor channels
will be at most 1\% of $ B_K $ and so much smaller than the statical
and other systematic errors at $ \beta = 5.7 $ 
($ a^{-1} = 2 GeV $).
This gives us a great confidence in our approximate matching at
one loop level, where such terms of wrong flavor channels, 
which enters at order $ g^2 $, are neglected.
From Figures \ref{fig:9} and \ref{fig:10},
we also observe that the nearest neighboring data points
have stronger correlation (less fluctuating) 
in the cubic wall source than
in the even-odd wall source.
The four times more floating point computation
in the cubic wall source than in the even-odd wall source
explains why the error bars of the data in Figure \ref{fig:10} is about half of those
in Figure \ref{fig:9}.
Hence, we conclude that the even-odd wall source is equivalent to
the cubic wall source from the standpoint of statistics.
We believe that it is better to
use the cubic wall source in weaker
coupling simulations because the unwanted 
contamination from the degenerate and excited
hadronic states becomes even more severe there.
\section{Data Interpretation}
Here, we interpret the numerical results
in terms of the physics.
First of all, we will address the technical questions
about the improved wall source methods.
Second, we will compare 
the numerical results of the one spin trace
form and the two spin trace form and discuss the meaning
of the consistency between the two formalism.
Third, we will discuss the chiral behavior of $ B_K $,
and the chiral behavior of the individual components
making up $ B_K $ ({\em i.e.} $ B_A $, $ B_{A1} $, $ B_{A2} $,
$ B_V $, $ B_{V1} $, $ B_{V2} $).
In addition, we will
discuss the effects of non-degenerate quark anti-quark
pairs on the chiral behavior of these quantities. 
We will also discuss
the chiral behavior from the standpoint of both  $1/N_c$
suppression and (partially quenched) chiral perturbation
theory. 
Fourth, we will compare our numerical results with
earlier works and discuss the effect of quenching
on $ B_K $ measurements. 
Finally, we will present our best
value of $ B_K $ in the physical limit as well as
our fitting procedure.
\subsection{Comparison of Wall Sources}
\begin{figure}[t]

\centering

\epsfxsize=\hsize\epsfbox{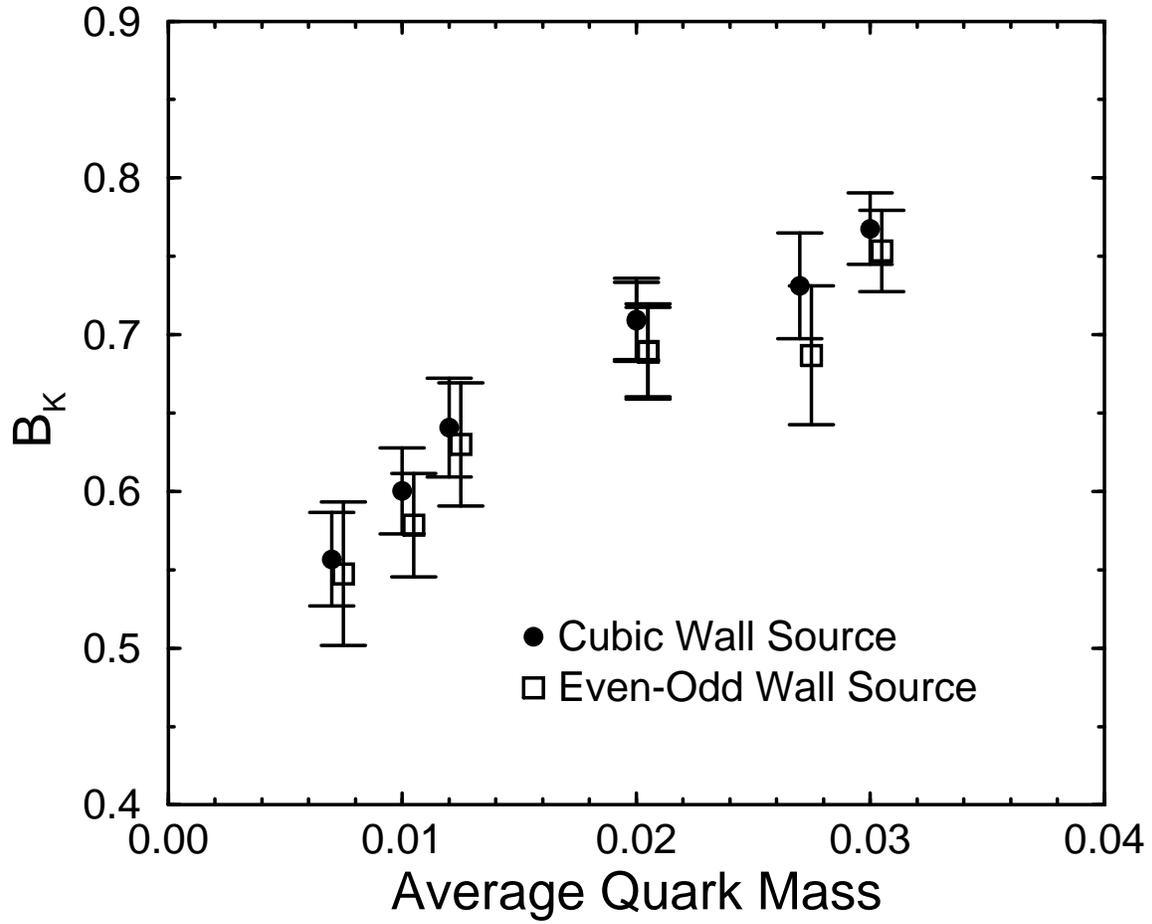}

\caption{Unrenormalized $ B_K $ with respect to average quark mass.
The filled circles represent the data from
the cubic wall source. The empty squares
represent the data from the even-odd wall
source. All the data are obtained using
the two spin trace formalism.} 

\label{fig:11}

\end{figure}

Numerical results for unrenormalized (naive, bare,
or tree-level) $ B_K $ for various average quark mass, 
calculated both by the even-odd wall source and by the cubic wall source
are shown in Figure \ref{fig:11}. 
The values of $ B_K $
calculated by the two wall sources
agree within errors.
The error bars for $ B_K $ calculated by the
cubic wall source are about half of those of 
the even-odd wall source.
The cubic wall source method
takes four times more computational time than the
even-odd wall source method.
We conclude that for $ B_K $ measurements at $ \beta = 5.7 $
the cubic wall source results are consistent
with the even-odd wall source results.
\subsection{Comparison of One Spin Trace Form
and Two Spin Trace Form}
\begin{figure}[t]

\centering

\epsfxsize=\hsize\epsfbox{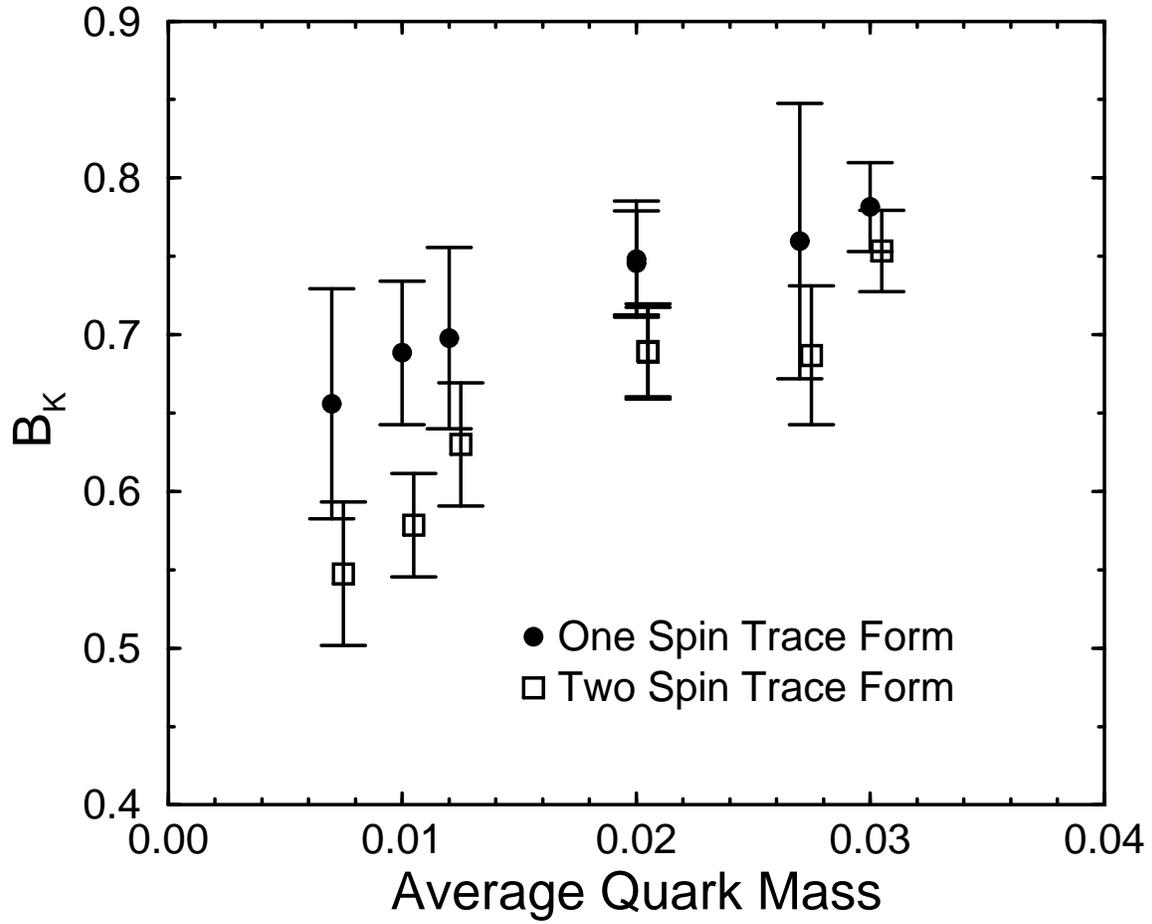}

\caption{Unrenormalized $ B_K $ with respect to average quark mass.
The filled circles represent the data from
the one spin trace form.
The empty squares
represent the data from the two
spin trace form.
Both data are obtained using
an even-odd wall source.} 

\label{fig:12}

\end{figure}

\begin{figure}[t]
 
\centering

\epsfxsize=\hsize\epsfbox{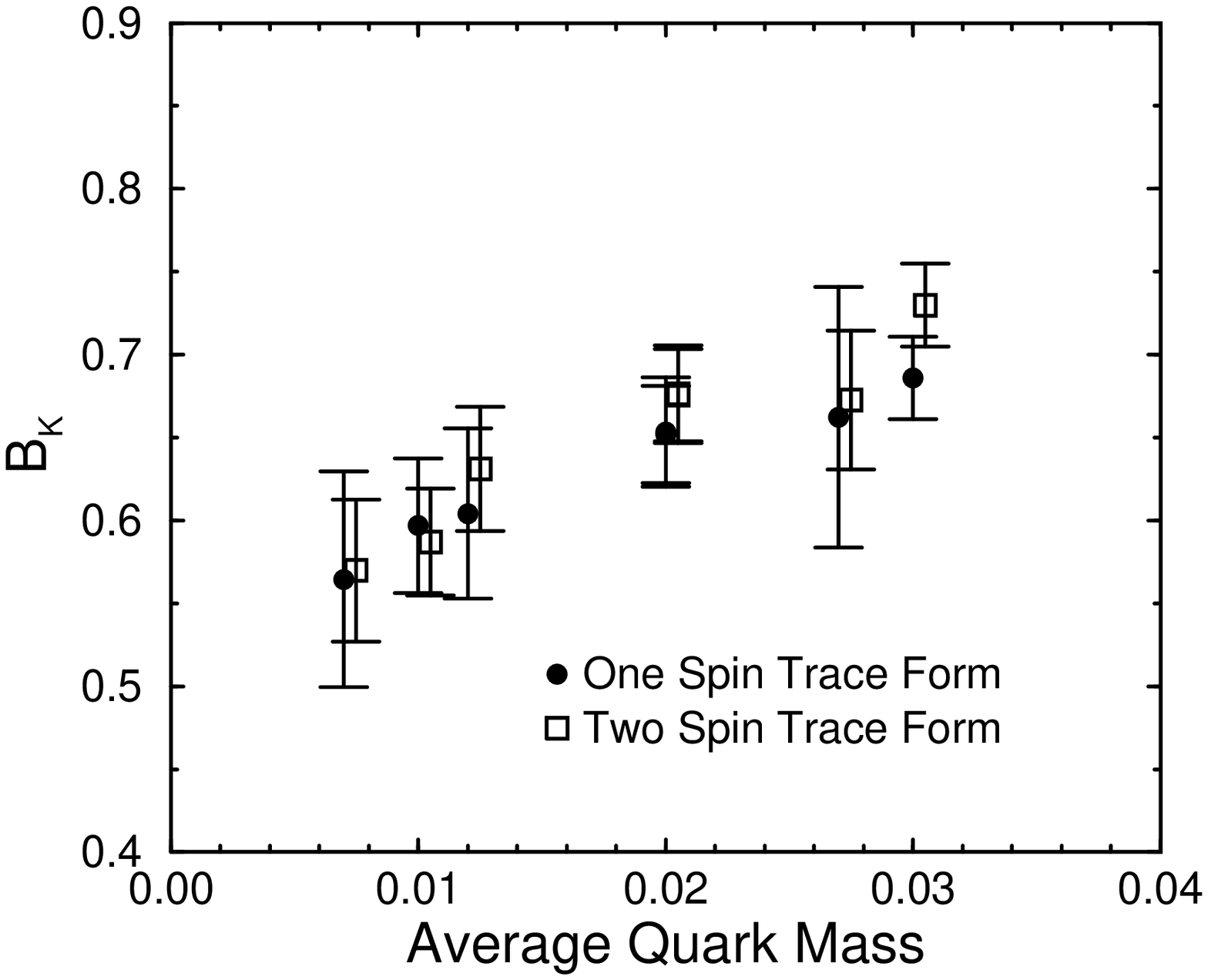}

\caption{Tadpole-improved renormalized
$ B_K $ with respect to average quark mass.
$ \mu = \pi/a $.
The filled circles represent the results from
the one spin trace form.
The empty squares
represent the results from the two
spin trace form.
Both results are obtained  using  
an even-odd wall source.}

\label{fig:13}

\end{figure}

\begin{figure}[t]
 
\centering

\epsfxsize=\hsize\epsfbox{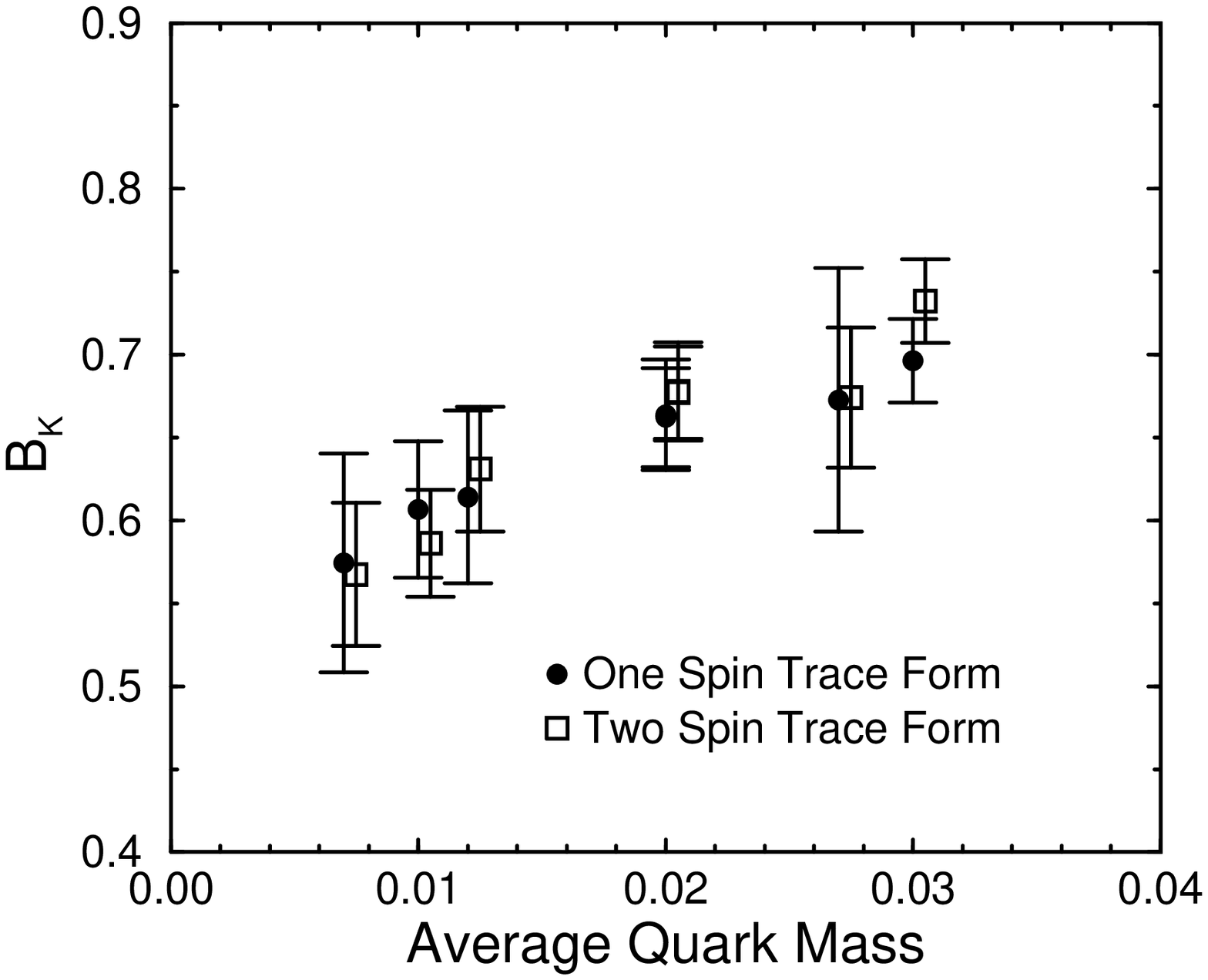}

\caption{RG improved ($ \overline{MS} $) renormalized
$ B_K $ with respect to average quark mass.
$ \mu = \pi/a $.
The filled circles represent the results from
the one spin trace form.
The empty squares
represent the results from the two
spin trace form.
Both results are obtained  using  
an even-odd wall source.}

\label{fig:14}
 
\end{figure}

The numerical results for unrenormalized 
$ B_K $ in both one spin trace and two spin trace 
methods are presented in Figure \ref{fig:12}.
The numerical results for one-loop renormalized 
(tadpole improved through mean field theory 
\cite{wlee0,sharpe3,sharpe4,japan1,parisi,lepage})
$ B_K $ both in the one spin trace
and two spin trace methods are shown in Figure \ref{fig:13}.
The numerical results for one-loop
renormalized (RG improved \cite{wlee0,wlee1}) $ B_K $
both in the one spin trace 
and two spin trace form are
given in Figure \ref{fig:14}.
From Figures \ref{fig:12}, \ref{fig:13} and \ref{fig:14},
we notice that 
the results for the renormalized $ B_K $ calculated 
in the one spin trace method  agree with those in the
two spin trace method better than 
those for the unrenormalized $ B_K $.
Let us explain how we have obtained the renormalized coupling
constant for the perturbative expansion.
The one-loop renormalization of the $ \Delta S = 2 $
four-fermion operators on the lattice
is explained in Ref. \cite{wlee0,sharpe4,japan1}
in detail. 
The detailed explanation of matching between
the continuum and lattice observables at one loop level
is also given in Ref. \cite{wlee0,sharpe4,japan1}.
We discuss here how to obtain
the coupling constant for the renormalization
of the composite operators.
For perturbative matching at one-loop level,
one needs a well-defined coupling constant to use as
the perturbative expansion parameter \cite{lepage,wlee1}.
We have chosen the $ \overline{MS} $
coupling constant at $ \mu_{\overline{MS}} = \pi/a $ scale
as our perturbative expansion parameter
\cite{japan0}.
There are two methods to obtain the $  \overline{MS} $
coupling constant from the bare lattice coupling constant.
One is a non-perturbative approach using tadpole improvement
by mean field theory \cite{lepage} 
and the other is a purely perturbative
approach using renormalization group improvements \cite{wlee1}.
We will refer to these two approaches to obtaining
$ g^2_{\overline{MS}} $ as the ``tadpole improved" and 
``RG improved" methods.
We use both methods to obtain the $ \overline{MS} $
coupling constant at the renormalization scale $ \pi/a $.

The details of the non-perturbative approach
of tadpole improvement by mean field theory is given
in Ref \cite{lepage,wlee0,wlee1,sharpe4,japan1}.
The $ \overline{MS} $
coupling constant ($ g^2_{\overline{MS}} $)
is related to the tadpole-improved coupling constant 
($  g^2_{MF} $) and the bare lattice coupling constant
($ g^2_0 $) through mean field theory \cite{wlee0,wlee1}:
\begin{eqnarray} 
& & g^2_{MF} \equiv \frac{g^2_0(a)}
{ {\rm Re} \langle \frac{1}{3} {\rm Tr} U_{\Box} \rangle}
\\
& & g^2_{\overline{MS}}(\mu_{\overline{MS}}) =
 g^2_{MF}\left[ 1 - \beta_0 g^2_{MF}
\left\{ 2 \ln (a \mu_{\overline{MS}}) + 2 \ln\left(
\frac{\Lambda_{\rm Latt}}{\Lambda_{\overline{MS}}}
\right) \right\}
\right.
\nonumber \\
& & \hspace{30mm} 
\left.  -\frac{1}{3}  g^2_{MF} +
O(g^4_{MF})\right] 
\end{eqnarray}
where $ g^2_0(a) $ is the bare lattice  coupling constant,
$ U_{\Box} $ is a unit gauge
link plaquette,
and 
\begin{eqnarray}
\beta_0 = \frac{11}{16 \pi^2}
\left( 1 - \frac{ 2 N_f}{33} \right) \ .
\end{eqnarray}
For $ \beta = 5.7 $ and $ N_f = 2 $,
we obtain the tadpole improved $ \overline{MS} $
coupling constant:
\begin{eqnarray}
& & g^2_{MF} = 1.82
\nonumber \\
& & g^2_{\overline{MS}}\left( \frac{\pi}{a} \right)
= 1.78 \ .
\label{tadpole-g}
\end{eqnarray}
The coupling constant in Eq. (\ref{tadpole-g})
is used for tadpole improved matching
between the continuum and the lattice observables.
The details about a perturbative approach of RG improvement
is explained in Ref. \cite{wlee1}.
In this approach,
the $ \overline{MS} $ coupling constant 
($ g^2_{\overline{MS}} $) is
related to the bare lattice coupling constant
($ g^2_0(a) $):
\begin{eqnarray}
& & g^2_{\overline{MS}} ( \mu_{\overline{MS}}) = 
\frac{g^2_0(a)}
{ 1 - t \beta_0 g^2_0(a) }
\\
& &\mbox{where } \ 
t = -2 \left[  \ln(a\mu_{\overline{MS}}) + 
\ln \left( \frac{\Lambda_{\rm Latt}}
{\Lambda_{\overline{MS}}} \right) \right] \ ,
\nonumber
\end{eqnarray}
For $ \beta = 5.7 $ and $ N_f = 2 $,
we obtain the $ \overline{MS} $ coupling constant
at $ \mu_{\overline{MS}} = \frac{\pi}{a} $ scale:
\begin{eqnarray}
& & g^2_0(a) = 1.05263
\nonumber \\
& & g^2_{\overline{MS}} \left( \frac{\pi}{a} \right)  
= 1.61 \ .
\label{per-g}
\end{eqnarray}
The coupling constant in Eq. (\ref{per-g})
is used for RG improved matching
between the continuum and the lattice observables.
\subsection{Chiral Logarithms and  Non-degenerate Quark Mass Pairs}
\begin{figure}[t]

\centering

\epsfxsize=\hsize\epsfbox{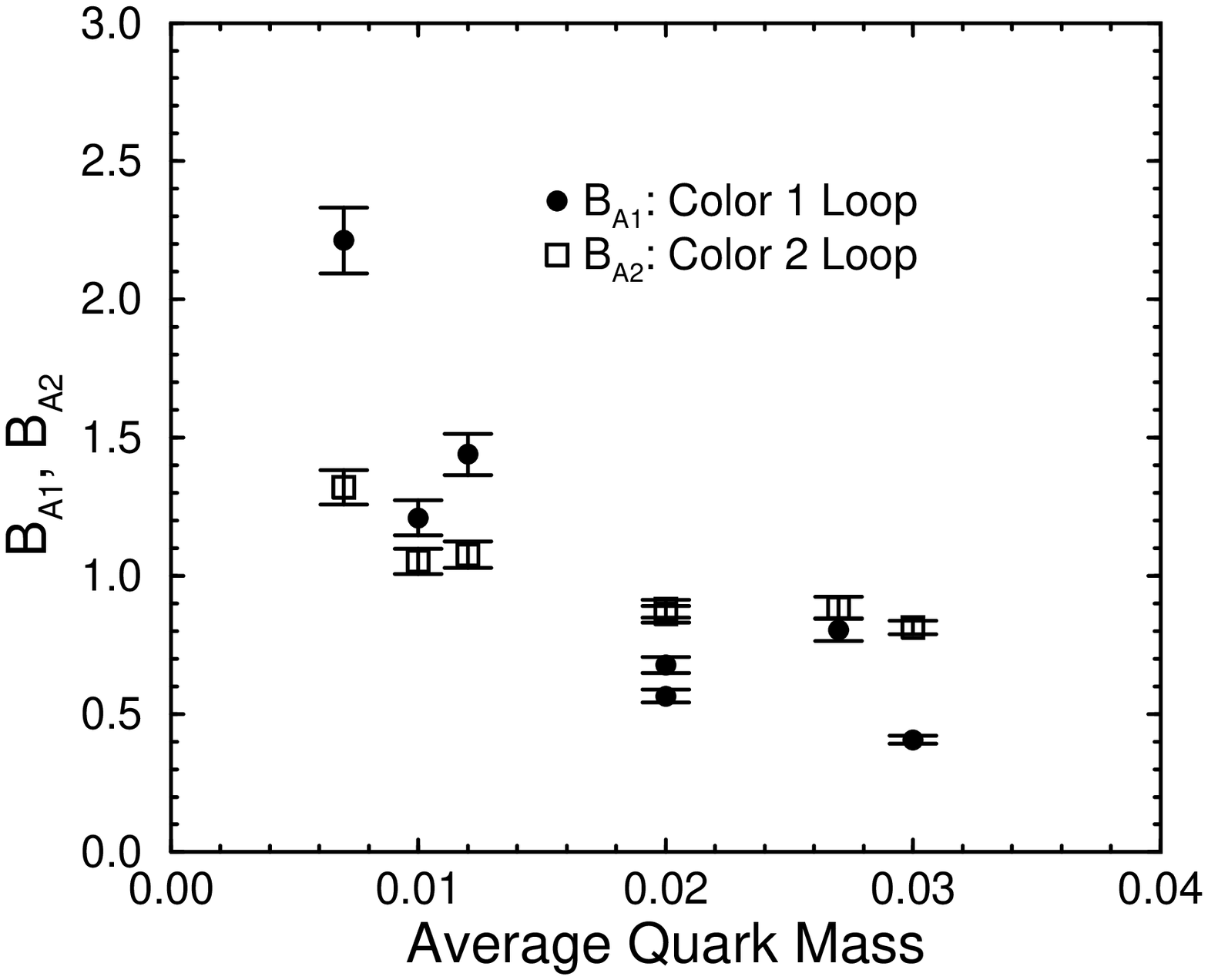}

\caption{$ B_{A1} $, $ B_{A2} $ with respect to average quark mass.
Three data points with average quark mass $ \in \{0.01,
0.02, 0.03\} $ correspond to the degenerate quark masses.
The other four data points
correspond to the non-degenerate masses.
The data are obtained using
the two spin trace formalism with the cubic wall
source method.}

\label{fig:15}

\end{figure}

\begin{figure}[t]
 
\centering

\epsfxsize=\hsize\epsfbox{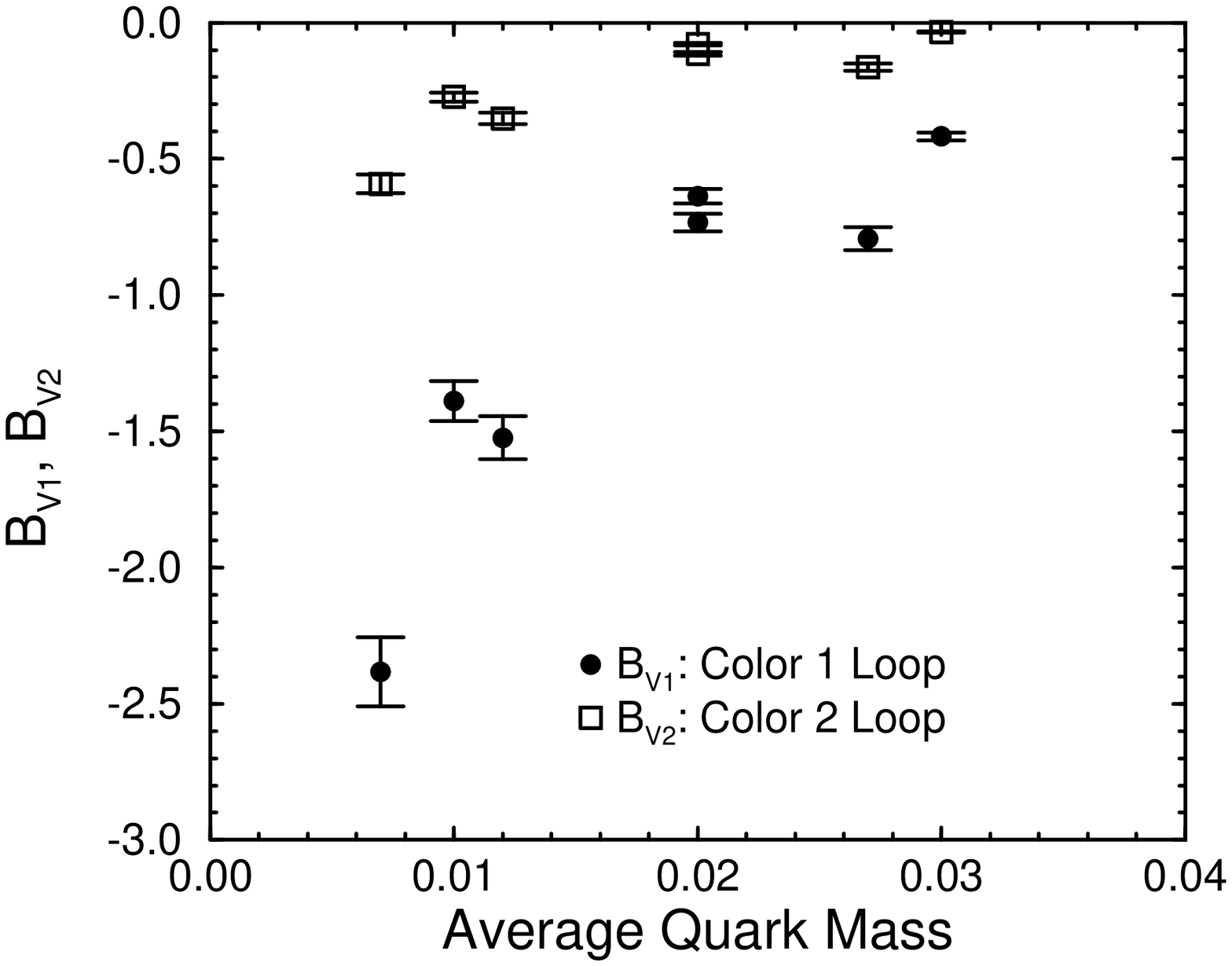}

\caption{$ B_{V1} $, $ B_{V2} $ with respect to average quark mass.
Three data points with average quark mass $ \in \{0.01,
0.02, 0.03\} $ correspond to the degenerate quark masses.
The other four data points
correspond to the non-degenerate masses.
The data are obtained using
the two spin trace formalism with the cubic wall
source method.}

\label{fig:16}

\end{figure}

\begin{figure}[t]
 
\centering

\epsfxsize=\hsize\epsfbox{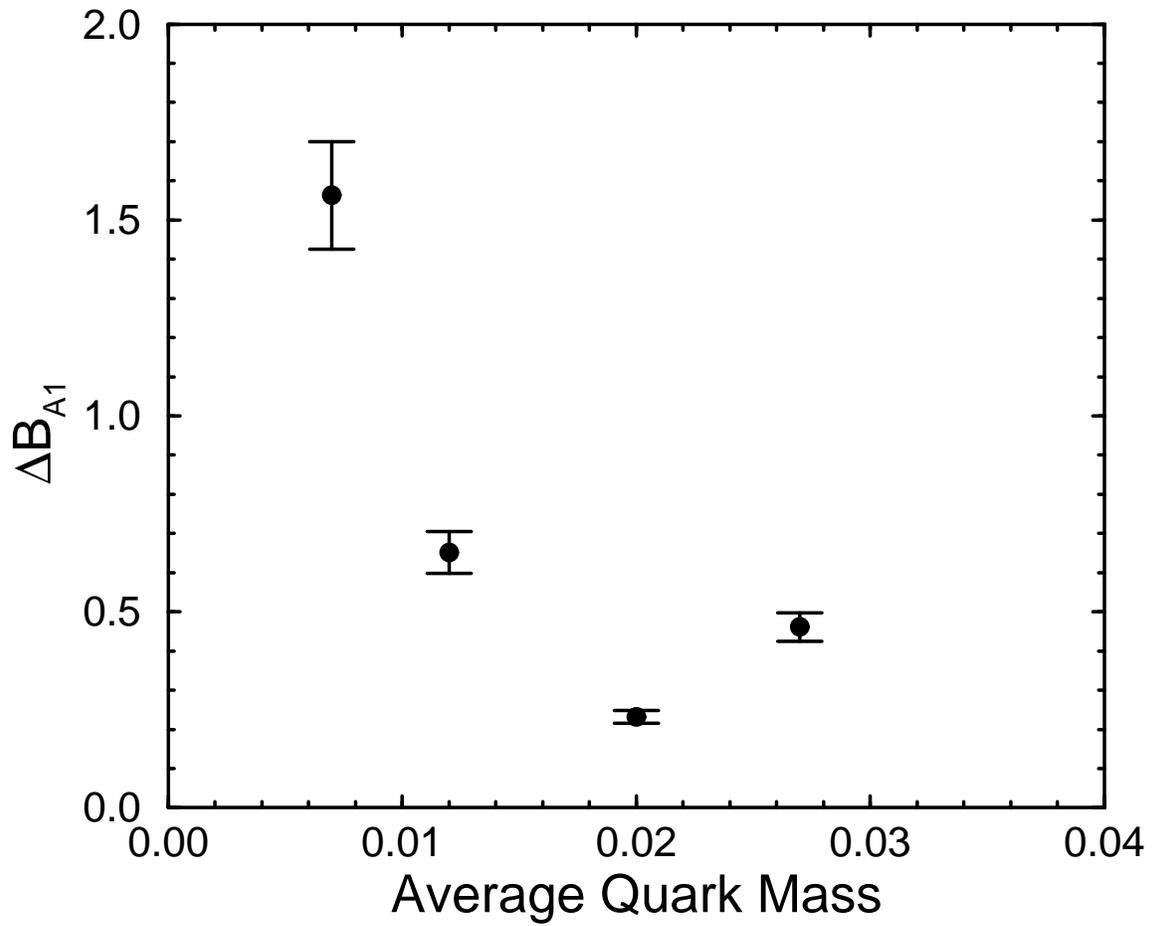}

\caption{
$ \Delta B_{A1} $ with respect to average quark mass.
The quark mass pairs are (0.004, 0.01), (0.004, 0.02),
(0.01, 0.03) and (0.004, 0.05).
The data are obtained using
the two spin trace formalism with the cubic wall
source method.}

\label{fig:17}

\end{figure}

There have been a number of theoretical
attempts to understand
the chiral behavior of the $ K $ meson $ B $ parameters
in terms of chiral perturbation theory
\cite{sharpe30,sonoda0,donoghue,bardeen}.
In order to discuss the chiral behavior of the
$ B $ parameters in an organized way,
we need to consider a theory with four valence flavors: 
$ S $ and $ S'$
both with mass $ m_s $ as well as $ D $ and $ D' $, 
both with mass $ m_d $.
Let the $ K^0$ be the $ \bar{S} \gamma_5 D $ pion and the
$ K'^0 $ be the corresponding state with
primed quarks ({\em i.e.} $ \bar{S}'\gamma_5 D' $ pion).
Let us define $ B_{V1} $, $ B_{V2} $, $ B_{A1} $
and $ B_{A2} $ as:
\begin{eqnarray}
B_{V1} & = & \frac{ \langle \bar{K}'^0 \mid
[\bar{S}'_{a} \gamma_{\mu} D'_{b} ]
[\bar{S}_{b} \gamma_{\mu} D_a ] \mid K^0 \rangle }
{ \frac{4}{3} \langle \bar{K}'^0  \mid
\bar{S}'_{a} \gamma_{\mu} \gamma_5 D'_{a} \mid 0 \rangle
\langle 0 \mid 
\bar{S}_b \gamma_{\mu} \gamma_5 D_b  \mid K^0 \rangle}
\\
B_{V2} & = & \frac{ \langle \bar{K}'^0 \mid
[\bar{S}'_{a} \gamma_{\mu} D'_{a} ]
[\bar{S}_{b} \gamma_{\mu} D_b ] \mid K^0 \rangle }
{ \frac{4}{3} \langle \bar{K}'^0  \mid
\bar{S}'_{a} \gamma_{\mu} \gamma_5 D'_{a} \mid 0 \rangle
\langle 0 \mid
\bar{S}_b \gamma_{\mu} \gamma_5 D_b  \mid K^0 \rangle} 
\\
B_{A1} & = & \frac{ \langle \bar{K}'^0 \mid
[\bar{S}'_{a} \gamma_{\mu} \gamma_5 D'_{b} ]
[\bar{S}_{b} \gamma_{\mu} \gamma_5 D_a ] \mid K^0 \rangle }
{ \frac{4}{3} \langle \bar{K}'^0  \mid
\bar{S}'_{a} \gamma_{\mu} \gamma_5 D'_{a} \mid 0 \rangle
\langle 0 \mid
\bar{S}_b \gamma_{\mu} \gamma_5 D_b  \mid K^0 \rangle}
\\
B_{A2} & = & \frac{ \langle \bar{K}'^0 \mid
[\bar{S}'_{a} \gamma_{\mu} \gamma_5 D'_{a} ]
[\bar{S}_{b} \gamma_{\mu} \gamma_5 D_b ] \mid K^0 \rangle }
{ \frac{4}{3} \langle \bar{K}'^0  \mid
\bar{S}'_{a} \gamma_{\mu} \gamma_5 D'_{a} \mid 0 \rangle
\langle 0 \mid
\bar{S}_b \gamma_{\mu} \gamma_5 D_b  \mid K^0 \rangle}
\end{eqnarray}
where $ a $ and $ b $ are color indices.
The $ B_A $, $ B_V $ and $ B_K $  are expressed in terms
of $ B_{V1} $, $ B_{V2} $, $ B_{A1} $ and $ B_{A2} $:
\begin{eqnarray}
B_A & =  & B_{A1} + B_{A2} \\
B_V & =  & B_{V1} + B_{V2} \\
B_K & =  & B_V + B_A 
\end{eqnarray}
Let us discuss
the chiral behavior of the B parameters defined above
($ B_{A1}$, $ B_{A2} $, $ B_{V1}$, $ B_{V2} $).
In the vacuum saturation approximation with
$ 1/N_C $ suppression, $ B_{A1}$ = 0.25,
$ B_{A2} $ = 0.75 and $ B_{V1} = B_{V2} $ = 0
(obviously wrong!).
For $ m_u = m_d \ne m_s $, chiral perturbation
theory predicts the following results \cite{sharpe1,sharpe30}:
\begin{eqnarray}
\label{b-v1}
B_{V1} & = &
\frac{3}{8}
\left[( \alpha_{-} - \delta_1)
+ \alpha_{+} A
+ \beta_1 \frac{\mu^2}{ m_k^2} C
+ \gamma_1 D
\right]
\\
\label{b-v2}
B_{V2} & = &
\frac{3}{8}
\left[( \alpha_{+} - \delta_2)
+ \alpha_{-} A
+ \beta_2 \frac{\mu^2}{ m_k^2} C
+ \gamma_2 D
\right]
\\
\label{b-a1}
B_{A1} & = &
\frac{3}{8}
\left[( \alpha_{-} + \delta_1)
+ \alpha_{+} A
- \beta_1 \frac{\mu^2}{ m_k^2} C
- \gamma_1 D
\right]
\\
\label{b-a2}
B_{A2} & = &
\frac{3}{8}
\left[( \alpha_{+} + \delta_2)
+ \alpha_{-} A
- \beta_2 \frac{\mu^2}{ m_k^2} C
- \gamma_2 D
\right]
\end{eqnarray}
where
in the vacuum saturation approximation
(tree-level chiral perturbation),
$ \alpha_+ = 1 $, $ \alpha_- = \frac{1}{3} $,
$ \delta_1 = \frac{1}{3} $, $ \delta_2 = 1 $,
$ \beta_1 = 1 $, $ \beta_2 = \frac{1}{3} $ while 
$ \gamma_1 $ and $ \gamma_2 $  is not determined.
$ A $, $ C $ and $ D $ are defined as
\begin{eqnarray}
A & = & I_2(m_K) - \frac{m_K^2 + m_{ss}^2}{2 m_K^2}
I_1(m_{ss}) - \frac{m_K^2 + m_{dd}^2}{2 m_K^2}
I_1(m_{dd})
\\
C & = & I_1(m_{ss}) + I_1(m_{dd}) - 2 I_1(m_K)
- 2 I_2(m_K)
\\
D & = & 2 I_1(m_K) - I_2(m_K) 
- \frac{m_K^2 + m_{ss}^2}{2 m_K^2} I_1(m_{ss})
\nonumber \\
& &
\hspace{20mm}
- \frac{m_K^2 + m_{dd}^2}{2 m_K^2} I_1(m_{dd})
\end{eqnarray}
where
\begin{eqnarray}
I_1(m) & \equiv & 
\frac{1}{f^2} \int^{\Lambda} \frac{d^4k}{(2\pi)^4}
\frac{1}{k^2 + m^2}
\nonumber \\
& = & \frac{\Lambda^2}{(4\pi f)^2}
+ \left( \frac{m}{4\pi f} \right)^2 
\ln\left( \frac{ m^2 }{\Lambda^2} \right)
+ O\left( \frac{ m^2 }{\Lambda^2} \right)
\\
I_2(m) & \equiv &
\frac{ m^2 }{ f^2 }
\int^{\Lambda} \frac{d^4k}{(2\pi)^4}
\left( \frac{1}{k^2 + m^2} \right)^2
\nonumber \\
& = & - \left( \frac{m}{4\pi f} \right)^2
\ln\left( \frac{ m^2 }{\Lambda^2} \right)
- \frac{m^2}{(4\pi f)^2}
+ \frac{m^2}{(4\pi f)^2}
O\left( \frac{ m^2 }{\Lambda^2} \right) \ .
\end{eqnarray}
$ \Lambda $ is introduced as a momentum cut-off
regularization for chiral perturbation theory.
$ m_{ss}^2 = 2 \mu m_s $,
$ m_{dd}^2 = 2 \mu m_d $ and 
$ m_K^2 = \mu (m_s + m_d) $ ({\em i.e.} 
$ 2 m_K^2 = m_{ss}^2 + m_{dd}^2 $).
Note there is no quadratic divergence in $ B_{V1} $,
$ B_{V2} $, $ B_{A1} $, and $ B_{A2} $ even though
$ I_1(m) $ has a quadratic divergence.
The quadratic divergences in $ C $ and $ D $ cancel out.
The quadratic divergences in $ A $ is not a function of 
quark masses $ m_s $ or $ m_d $.
Hence, these quadratic divergences in $ A $ can be absorbed
into the coefficients $ \alpha_- $ and $ \alpha_+ $.
As a summary of chiral perturbation theory,
let us choose $ \Lambda = 4 \pi f $ and
rewrite the results for $ B_{V1} $,
$ B_{V2} $, $ B_{A1} $, and $ B_{A2} $
as follows:
\begin{eqnarray}
B_{V1} & = & \frac{3}{8} (2 \beta_1) \frac{\mu^2}{(4 \pi f)^2}
\ln (z) + \frac{3}{8} ( \alpha_- - \delta_1 )
\nonumber \\
& &
+ \frac{3}{8} \beta_1 \frac{\mu^2}{(4 \pi f)^2}
\left[ \ln (1- \epsilon^2) 
+ \epsilon \ln \left( \frac{1+\epsilon}{1-\epsilon} \right) \right]
\nonumber \\
& &
+ ( \sigma_{\scriptscriptstyle V1} + 
\eta_{\scriptscriptstyle V1} \epsilon^2) z \ln(z)
+ ( \xi_{\scriptscriptstyle V1} + \rho_{\scriptscriptstyle V1} \epsilon^2) z 
+ {\cal O}(\epsilon^4 z \ln (z) )
\label{bk-v1}
\\
B_{V2} & = & \frac{3}{8} (2 \beta_2) \frac{\mu^2}{(4 \pi f)^2}
\ln ( z)
+ \frac{3}{8} ( \alpha_+ - \delta_2 )
\nonumber \\
& &
+ \frac{3}{8} \beta_2 \frac{\mu^2}{(4 \pi f)^2}
\left[ \ln (1- \epsilon^2) 
+ \epsilon \ln \left( \frac{1+\epsilon}{1-\epsilon} \right) \right]
\nonumber \\
& &
+ ( \sigma_{\scriptscriptstyle V2} + 
\eta_{\scriptscriptstyle V2} \epsilon^2) z \ln(z)
+ ( \xi_{\scriptscriptstyle V2} + \rho_{\scriptscriptstyle V2} \epsilon^2) z 
+ {\cal O}(\epsilon^4 z \ln (z) )
\label{bk-v2}
\\
B_{A1} & = & - \frac{3}{8} (2 \beta_1) \frac{\mu^2}{(4 \pi f)^2}
\ln ( z)
+ \frac{3}{8} ( \alpha_- + \delta_1 )
\nonumber \\
& &
- \frac{3}{8} \beta_1 \frac{\mu^2}{(4 \pi f)^2}
\left[ \ln (1- \epsilon^2) 
+ \epsilon \ln \left( \frac{1+\epsilon}{1-\epsilon} \right) \right]
\nonumber \\
& &
+ ( \sigma_{\scriptscriptstyle A1} + 
\eta_{\scriptscriptstyle A1} \epsilon^2) z \ln(z)
+ ( \xi_{\scriptscriptstyle A1} + \rho_{\scriptscriptstyle A1} \epsilon^2) z 
+ {\cal O}(\epsilon^4 z \ln (z) )
\label{bk-a1}
\\
B_{A2} & = & - \frac{3}{8} ( 2 \beta_2 ) \frac{\mu^2}{(4 \pi f)^2}
\ln ( z)
+ \frac{3}{8} ( \alpha_+ + \delta_2 )
\nonumber \\
& &
- \frac{3}{8} \beta_2 \frac{\mu^2}{(4 \pi f)^2}
\left[ \ln (1- \epsilon^2) 
+ \epsilon \ln \left( \frac{1+\epsilon}{1-\epsilon} \right) \right]
\nonumber \\
& &
+ ( \sigma_{\scriptscriptstyle A2} + 
\eta_{\scriptscriptstyle A2} \epsilon^2) z \ln(z)
+ ( \xi_{\scriptscriptstyle A2} + \rho_{\scriptscriptstyle A2} \epsilon^2) z 
+ {\cal O}(\epsilon^4 z \ln (z) )
\label{bk-a2}
\end{eqnarray}
where the constant terms proportional to $ \beta_i $ are
absorbed into $ \delta_i $, 
and
\begin{eqnarray}
z & \equiv & \frac{m^2_K}{ (4 \pi f)^2 }
\\
\epsilon & \equiv & \frac{ m_s - m_d} { m_s + m_d } \ .
\end{eqnarray}
Here we keep full functional form with respect to $ \epsilon $
for theose terms of order $ \ln(z) $ and $ z^0 $.
For those terms of order $ z\ln(z) $ and $ z $, we do 
Taylor expansion with respect to $ \epsilon $ and keep
only the terms of order $ \epsilon^0 $ and $ \epsilon^2 $.
The coefficients
$ \sigma_i $, $ \eta_i $, $ \xi_i $ and $ \rho_i $ 
($ i \in \{V1, V2, A1,A2 \} $) are unknown. 
All the results in Eqn. (\ref{b-v1}-\ref{b-a2}) and
Eqn. (\ref{bk-v1}-\ref{bk-a2})
are calculated using full QCD as
the fundamental theory.
Note, $ B_{V1} $, $ B_{V2} $, $ B_{A1} $, and $ B_{A2} $
have a branch point at $ \epsilon = \pm 1 $ which is non-singular.
Let us set the domain of $ \epsilon $
to $ -1 \le \epsilon \le 1 $ and $ \epsilon \in R $.
$  B_{V1} $, $ B_{V2} $, $ B_{A1} $, and $ B_{A2} $
have no singularity on this physical domain of $ \epsilon $.
Also note, $ B_{V1} $, $ B_{V2} $, $ B_{A1} $, and $ B_{A2} $
are even functions of $ \epsilon $ which means that
although we switch $ m_s $ with $ m_d $, the physics
does not know it at all ({\em i.e.} it does not change).
Hence, the power series expansion of $ B_{V1} $, $ B_{V2} $, $ B_{A1} $,
and $ B_{A2} $ with respect to $ \epsilon $ should have only
even powers of $ \epsilon $. 
The one-loop corrections to $ B_{V1} $, $ B_{V2} $
$ B_{A1} $ and $ B_{A2} $ include a term of the order
$ I_2(m_K)/ m_K^2 $ which is proportional to
$ \beta_i $ in Eqn. (\ref{b-v1}-\ref{b-a2}).
$ I_2(m_K)/ m_K^2 $ is proportional to $ \ln (m_K) $
which is logarithmically divergent 
as $ m_K \rightarrow 0 $. 
These logarithmically divergent terms
are called {\em enhanced chiral logarithms} \cite{sharpe30},
and were originally noticed 
by Langacker and Pagels \cite{langacker}. 
It is important to note that in full QCD,
the enhanced chiral logarithms
in Eqn. (\ref{b-v1}-\ref{b-a2}) and Eqn. (\ref{bk-v1}-\ref{bk-a2})
are all proportional to
$ \beta_i \frac{\mu^2}{m_K^2} I_2(m_K) $, which
depends only on the average mass of the quark antiquark
pair.
Let us adapt the above results to the (partially) quenched
limit of QCD \cite{sharpe30}.
Here, quenched approximation means neglecting all the internal
fermion loops and keeping only pure gauge interactions
({\em i.e.} sea quark mass is infinitely heavy),
while the partially quenched approximation implies that
the sea quark mass of the internal fermion loops is different
from the valence quark mass.
Both partially quenched and quenched approximations
have additional infra-red problems which are absent
in full QCD, since the sea quark fermion determinant
can not regulate the infra-red pole singularity of the
valence quark propagator.
There are two important differences
between (partially) quenched and
full QCD.
The first difference is that the meson eigenstates
are not the same.
The second difference comes from
$ \eta' $ loops, which can not contribute in full QCD
simply because the $ \eta' $ is too heavy.
The diagrams of the hairpin type 
($ \eta' $ loops) present in
the (partially) quenched chiral perturbation
for $ B_K $
vanish in the limit of $ m_s = m_d $
\cite{sharpe30}. 
The logarithmically divergent terms
which come from the hairpin diagrams 
are called {\em quenched chiral logarithms}
\cite{bernard0,bernard1,sharpe30}. 
Here, the key point is that in (partially)
quenched QCD, the quenched chiral logarithms in the
B parameters, if present, must be a function of
both average quark mass ($ m_s + m_d $) and
mass difference ($ m_s - m_d $) 
of the quark anti-quark pair,
while the enhanced chiral logarithms common in both full
and (partially) quenched QCD are not a function of
quark mass difference 
but a function of only average quark mass.
The quantity $ B_{A1} $ in Eqn. (\ref{b-a1}) 
has a finite constant term which is
a factor of around 1/3 smaller than that in $ B_{A2} $ in
Eqn. (\ref{b-a2}), while
the enhanced logarithmic term in $ B_{A1} $
is around 3 times larger than  that in  $ B_{A2} $.
The enhanced logarithmic term in $ B_{V1} $ is
also around 3 times larger than  that in $ B_{V2} $.
Therefore, we have chosen $ B_{A1} $
as a useful measurement
adequate to observe both the enhanced chiral logarithms
and, if present,
the (partially) quenched chiral logarithms.
We plot $ B_{A1} $ and $ B_{A2} $ with respect to
the average quark masses in Figure \ref{fig:15} 
and $ B_{V1} $ and $ B_{V2} $
with respect to the quark masses in Figure \ref{fig:16}.
One can see a difference in the chiral behavior of
$  B_{A1} $ and $ B_{V1} $ 
between the case with degenerate quark anti-quark pairs
of  mass: \{0.01, 0.02, 0.03\}
and the situation with non-degenerate quark anti-quark pairs
with the mass pairs: \{(0.004, 0.01),
(0.004, 0.02), (0.004, 0.05), (0.01, 0.03)\}.
In order to more precisely see the effect of non-degenerate
quark anti-quark pairs on $ B_{A1} $, we must interpolate
in quark mass.
Toward this end, we fit the data of the degenerate quark 
anti-quark pairs to the following function:
\begin{eqnarray}
B_{A1}^{\rm degenerate} (m_K) =
A_1 \log \left( \frac{m_K^2}{(4\pi f_\pi)^2} \right)
+ A_2 + A_3  \frac{m_K^2}{(4\pi f_\pi)^2} 
\log \left( \frac{m_K^2}{(4\pi f_\pi)^2} \right)
\  ,
\label{b-a1-fit}
\end{eqnarray}
and, second, subtract this form for the degenerate case 
from the non-degenerate data as follows:
\begin{eqnarray}
\Delta B_{A1} (m_K) = \frac{m_s+m_d}{m_s-m_d}
\left(  B_{A1} (m_K, m_s-m_d) - B_{A1}^{\rm degenerate} (m_K)
\right) \  ,
\end{eqnarray}
where $  B_{A1} (m_K, m_s-m_d) $ is the numerical data
for the non-degenerate quark mass pairs.
The covariance fitting results of the degenerate data on
the jack-knifed ensembles are
is
\begin{eqnarray}
B_{A1}^{\rm degenerate} (m_K,\ a^{-1} = 2.0 {\rm GeV} ) & = &
- 0.683(47) \log \left( \frac{m_K^2}{(4\pi f_\pi)^2} \right)
+ 1.305(97)
\nonumber \\
& & \hspace*{-20mm} + 3.96(35) \frac{m_K^2}{(4\pi f_\pi)^2}
\log \left( \frac{m_K^2}{(4\pi f_\pi)^2} \right) 
\hspace*{5mm} ( \chi^2 = 8.1 \times 10^{-23} ) \\
B_{A1}^{\rm degenerate} (m_K,\ a^{-1} = 1.8 {\rm GeV} ) & = &
- 0.300(25) \log \left( \frac{m_K^2}{(4\pi f_\pi)^2} \right)
+ 2.46(20)
\nonumber \\
& & \hspace*{-20mm} + 6.34(56) \frac{m_K^2}{(4\pi f_\pi)^2}
\log \left( \frac{m_K^2}{(4\pi f_\pi)^2} \right)
\hspace*{5mm} ( \chi^2 = 8.1 \times 10^{-23} ) \ .
\end{eqnarray}
Using chiral perturbation theory, one can estimate
the value of $ A_1 $ in Eq. (\ref{b-a1-fit}).
From Eq. (\ref{bk-a1}), 
\begin{eqnarray}
A_1 = -\frac{3}{8} (2 \beta_1 ) \frac{\mu^2}{ (4 \pi f_\pi)^2 } \ ,
\end{eqnarray}
where $ \mu \equiv m_K^2/ ( m_s + m_d ) $.
Using the results of our lattice QCD simulation
($ \mu \cong 2.3 a^{-1} $),
this amplitude $ A_1 $  is written in terms of
a poorly known quantity  $ \beta_1 \sim 1 $:
$ A_1 \cong - 11.6 \beta_1 $.
Using the value of $ K $ meson mass and
strange quark mass in the particle data book
($ m_s \cong 100 \sim 300 MeV $ and 
$ m_s/m_d \cong 25 $),
one can also obtain $ A_1 \cong - (0.5 \sim 3.5) \beta_1 $
in terms of a poorly known quantity $ \beta_1 $, where
the range of values is chosen to reflect the large
uncertainty in the chiral perturbation theory prediction.
The numerical data for $ \Delta B_{A1} (m_K) $ with respect to the
average quark mass are plotted in Figure \ref{fig:17}.
From Figure \ref{fig:17}, it is obvious that
there is a strong sensitivity to
the non-degenerate quark mass pairs
({\em i.e.} $ \Delta B_{A1} (m_K) $ is 
not only a function of $ (m_s +m_d) $ but also
a function of $ (m_s - m_d) $).
This additional divergence may be related to
(partially) quenched chiral logarithms.
It could also come from finite volume dependence on
the lightest mass of the quark anti-quark pair.
Something else might cause this unexpected divergence.
The key point is that there is an additional
divergence which is visible in our numerical simulations.
At any rate, we would like to raise 
the following questions:
\begin{itemize}
	\item $ \eta' $ loop:
	Could the additional hairpin diagram in partially
	quenched chiral perturbation theory
	explain quantitatively the additional divergence 
	which is sensitive to the non-degenerate quark masses?
	In fact, it is known that the contribution
	from $ \eta'$ loops vanishes as $ m_s \rightarrow m_d $
	\cite{sharpe30,sharpe1}.
	\item finite volume effect:
	Could the finite volume effect of $ B_{A1} $
	or $ B_{V1} $ be sensitive to the
	non-degenerate quark masses?
	The small eigenvalues and their density
	in lattice QCD is regulated by the finite volume.
	Could these small eigenvalues be set up such that
	they are sensitive to the lightest mass of the
	non-degenerate quark anti-quark pair? 
	\item fermion determinant in partially quenched QCD:
	In partially quenched QCD,
	the sea quark mass is different from the
	valence quark mass. 
	In partially quenched QCD,
	the fermion determinant of the sea quark
	may suppress the coefficient of the
	logarithmic divergence more efficiently
	than the case of quenched QCD, which has no
	fermion determinants at all. 
	Could one see the much larger effect of
	the non-degenerate quark anti-quark pair
	on $ B_{A1} $ and $ B_{V1} $ in quenched QCD
	than in partially quenched QCD?
	\item scaling violation:
	The anomaly current in staggered fermion formulation
	$ \bar{\chi} \overline{ ( \gamma_{\mu} \gamma_5 \otimes
	I)} \chi $ is not a conserved current for finite lattice
	spacing. Hence, the corresponding pseudo-Goldstone
	pion $ \bar{\chi} \overline{ ( \gamma_{5} \otimes I ) } \chi $
	has a serious contamination of finite lattice spacing, which
	is supposed to vanish by flavor symmetry restoration
	in the continuum limit of
	$ a \rightarrow 0 $.
	This suggests that 
	even in quenched QCD,
	$ \eta' $ in staggered fermion formulation 
	for finite lattice spacing may be much heavier
	than pseudo Goldstone pion $ \bar{\chi} 
	\overline{ ( \gamma_{5} \otimes \xi_5 ) } \chi $ and 
	that its mass may have a scaling violation term 
	of order $ a $ to make matters worse.
	How large is the scaling violation of $ \eta' $
	mass as a function of quark mass?
	\item \begin{sloppypar}
	other possibility:
	Is there something else related to the systematics
	of the non-degenerate quark masses on the lattice?
	\end{sloppypar}
\end{itemize}
One important thing is that
this dependence on quark mass difference
amplifies the enchanced chiral logarithms rather than
reducing them. 
We see this in Figure \ref{fig:15}, 
\ref{fig:16} and \ref{fig:17}.

\begin{figure}[t]
 
\centering

\epsfxsize=\hsize\epsfbox{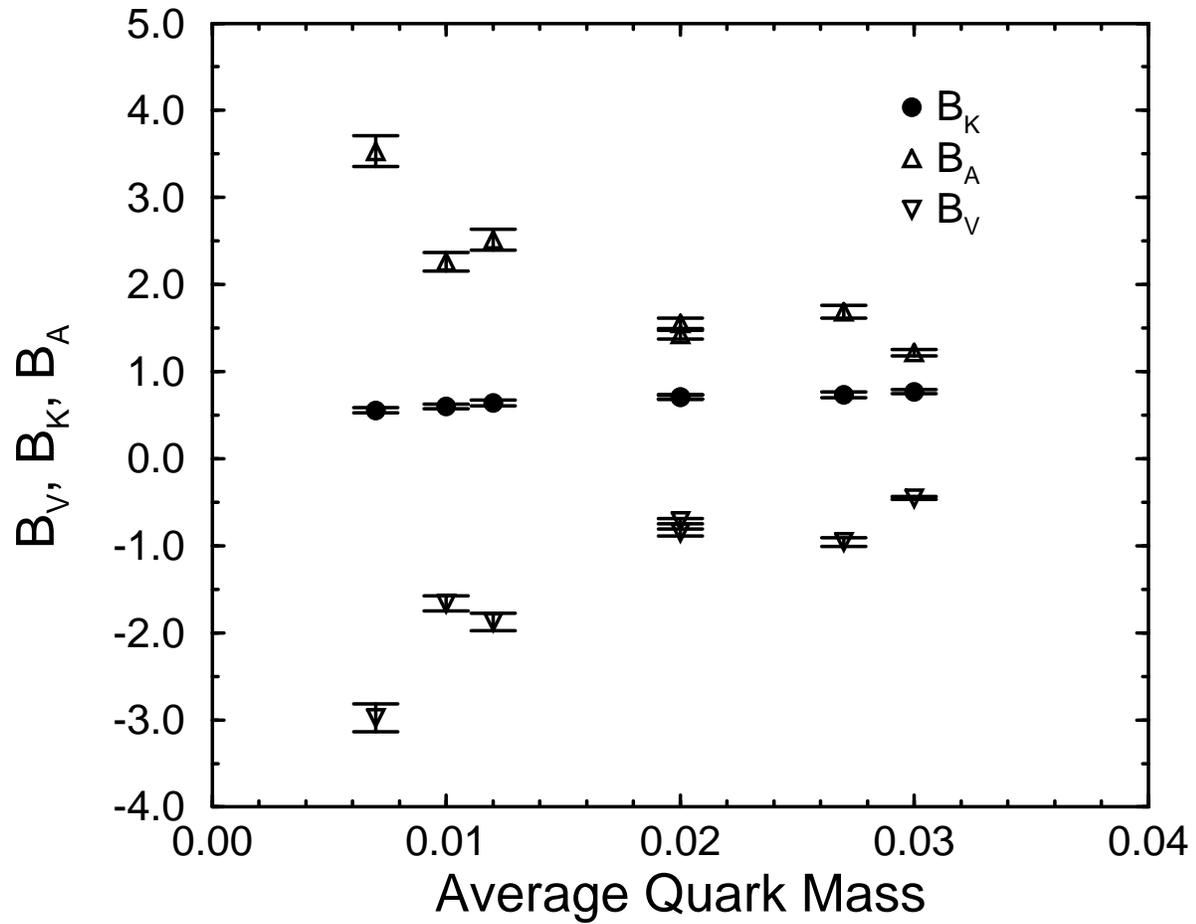}

\caption{Unrenormalized $ B_V $, $ B_K $ and $ B_A $ 
	with respect to average quark mass.
	The data are obtained using
	the two spin trace formalism with the cubic wall
	source method.}

\label{fig:18}

\end{figure}

$ B_A $ and $ B_V $ are plotted in Figure \ref{fig:18},
together with $ B_K $
with respect to the average quark masses.
One can notice that all these divergences related to
both (partially) quenched chiral logarithms (if present)
and enhanced chiral logarithms (present in
$ B_{A1} $, $ B_{A2} $, $ B_{V1} $ and $ B_{V2} $)
are canceled out in $ B_K $
which is finite in the chiral limit.
\subsection{Comparison with Earlier Work}
We now compare our numerical results of $ B_K $ with
those of other groups as well as comparing our
results at $ \beta = 5.7 $ (full QCD with $N_f = 2$)
with the results at $ \beta = 6.0 $ (Quenched QCD).
\begin{table}[t]
\renewcommand{\baselinestretch}{1}
\begin{center}
\small
\begin{tabular}{||c||c|c|c|c||} \hline \hline
 & \multicolumn{4}{c||}{Naive $ B_K $} \\
\cline{2-5}
$ m_{\rm valence} a $ & this work (JN) & this work (BS)  
& Kilcup & Ukawa {\em et al.} 
\\
\hline \hline
0.01  &  0.600(27) & 0.595(22) &  0.658(18) & 0.69(2) \\
\hline
0.02  &  0.709(25) & 0.727(24) &  0.771(11) & 0.75(1) \\
\hline
0.03  &  0.768(23) & 0.791(21) &  0.818(09) & 0.79(1) \\
\hline \hline
\end{tabular}
\caption{We compare our numerical results for
naive $ B_K $ (cubic wall source,
two spin trace form) with those of other groups
(Kilcup and Ukawa {\em et al.}).
$ \beta = 5.7 $. $ m_{\rm sea} a = 0.01 $.
JN implies that the errors are obtained through single-elimination
jack-knife method.
BS means that the data is analyzed by covariance fitting on the bootstrap
ensembles.}
\label{bk-compare}
\end{center}
\end{table}
There have been two groups to calculate
$ B_K $ at $ \beta = 5.7 $ 
with staggered fermions in full QCD.
Kilcup \cite{kilcup10} 
has calculated the unrenormalized (naive)
$ B_K $ at $ \beta = 5.7 $ ($ 16^3 \times 32 $,
full QCD with two dynamical flavors of a mass
$ m_{\rm sea} a $ = 0.01, 0.015, 0.025).
The lattice scale was $ a^{-1} = 1.9 \sim 2.0 $ GeV.
The number of independent
configurations was 50 and 
$ B_K $ measurements were done twice
in different locations on the lattice 
for each individual  configuration, 
to make the total number of configurations
equivalent to 100. 
Quark wall propagators were fixed in Landau gauge,
and  periodic boundary conditions in
space and Dirichlet boundary conditions in
the time direction were imposed.
\begin{table}[h]
\renewcommand{\baselinestretch}{1}
\begin{center}
\small
\begin{tabular}{||c||c|c||c|c||} \hline \hline
 & \multicolumn{2}{c||}{ $ B_{A1} $}
 & \multicolumn{2}{c||}{ $ B_{A2} $} \\
\cline{2-5}
$ m_{\rm valence} a $ & this work & Kilcup
& this work & Kilcup 
\\
\hline \hline
0.01  & 1.210(64) & 1.225(60) & 1.053(46) & 1.155(24) \\
\hline
0.02  & 0.565(23) & 0.575(15) & 0.860(30) & 0.932(13) \\
\hline
0.03  & 0.407(14) & 0.418(7) & 0.812(24) & 0.866(10) \\
\hline \hline
\end{tabular}
\caption{We compare our numerical results of
$ B_{A1} $  and $ B_{A2} $ (cubic wall source,
two spin trace form) with those of
Kilcup.
$ \beta = 5.7 $. $ m_{\rm sea} a = 0.01 $.
The errors are estimated through the standard
jack-knife procedure.}
\label{bk-compare-1}
\vspace*{5mm}
\begin{tabular}{||c||c|c||c|c||} \hline \hline
 & \multicolumn{2}{c||}{ $ B_{V1} $} &
   \multicolumn{2}{c||}{ $ B_{V2} $} \\
\cline{2-5}
$ m_{\rm valence} a $ & this work & Kilcup
& this work & Kilcup 
\\
\hline \hline
0.01  & - 1.389(73) & - 1.455(65) & - 0.273(17) & - 0.268(21)  \\
\hline
0.02  & - 0.638(26) & - 0.664(21) & - 0.0786(52) & - 0.0713(53) \\
\hline
0.03  & - 0.417(15) & - 0.435(12) & - 0.0342(22) & - 0.0302(21) \\
\hline \hline
\end{tabular}
\caption{We compare our numerical results of
$ B_{V1} $  and $ B_{V2} $ (cubic wall source,
two spin trace form) with those of 
Kilcup.
$ \beta = 5.7 $. $ m_{\rm sea} a = 0.01 $.
The errors are estimated through the standard
jack-knife procedure.}
\label{bk-compare-2}
\end{center}
\end{table}
Ukawa {\em et al.} \cite{japan0,japan20} 
studied  $ B_K $ at $ \beta = 5.7 $ on a lattice
of size $ 20^3 \times 20 $
(duplicated in the time direction)
with two flavors of dynamical staggered quarks of
mass $ m_{\rm sea} a = 0.01 $.
The lattice scale was $ a = 0.085 \sim 0.09 $ fm 
($ a^{-1} = 2.2 \sim 2.4 $ GeV).
Both Landau gauge operators and gauge-invariant
operators were used. 
Quark propagators were calculated with
Dirichlet (periodic) boundary condition
in the time (space) direction.
The differences between our numerical simulation
and those of other groups' are
the lattice size, the boundary conditions 
on the quark propagators in the time direction
and the color summation over the meson wall sources.
We are not summing the three values of the color index 
for the meson wall sources
on the individual configuration sample. 
Instead, we choose a different color index
for the meson wall sources in each measurement
(in other words, color indices of meson wall
sources are spread over
the configuration samples with equal statistical
weight instead of being summed
on each configuration
sample).
The $ B_K $ results of this work,
Kilcup and Ukawa {\em et al.}
are summarized in Table \ref{bk-compare}.
The $ B_{A1} $, $ B_{A2} $, $ B_{V1} $ and $ B_{V2} $
of this work and Kilcup \cite{kilcup40}
are compared in Tables \ref{bk-compare-1} and
\ref{bk-compare-2}.
We believe that the differences 
in Tables \ref{bk-compare}, \ref{bk-compare-1}
and \ref{bk-compare-2}
can be reasonably explained by the different lattice size,
the boundary conditions on the quark propagators,
color summation of the meson wall source, poor statistics,
the uncertainties in the lattice spacing,
{\em etc.}.
Hence, we conclude that all the measurements in Tables
\ref{bk-compare}, \ref{bk-compare-1} and \ref{bk-compare-2}
are consistent with one another.
It is also important to compare the results of full QCD
($ N_f = 2 $) with those of quenched QCD.
There have been three  groups to calculate 
$ B_K $ at $ \beta = 6.0 $ ($ a^{-1} = 2.0 \sim 2.1 $ GeV)
in quenched QCD. 
This corresponds to $ \beta = 5.7 $ in full QCD.
\begin{table}[t]
\renewcommand{\baselinestretch}{1}
\begin{center}
\small
\begin{tabular}{||c||c|c||c|c|c||} \hline \hline
 & \multicolumn{5}{c||}{Naive $ B_K $} \\
\cline{2-6}
$ m_{\rm valence} a $ &
\multicolumn{2}{c||}{full QCD ($ N_f = 2 $)} &
\multicolumn{3}{c||}{quenched QCD} \\
\cline{2-6}
& this work (JN) & this work (BS) 
& Sharpe {\em et al.} & Kilcup & Ukawa {\em et al.} \\
\hline \hline
0.01  &  0.600(27) & 0.595(22) & 0.68(2) & 0.697(29) & 0.69(2)  \\
\hline
0.02  &  0.709(25) & 0.727(24) & 0.73(1) & 0.749(16) & 0.74(1) \\
\hline
0.03  &  0.768(23) & 0.791(21) & 0.78(1) & 0.777(13) & 0.78(1) \\
\hline \hline
\end{tabular}
\caption{The full QCD calculation of naive (unrenormalized) 
$ B_K $ (cubic wall source, two spin trace form) is compared
with that of quenched QCD (Sharpe {\em et al.}, Kilcup and
the Ukawa {\em et al.} group). JN means that 
the errors are estimated through the standard jack-knife procedure.
BS means that the data is analyzed
by covariance fitting on the bootstrap ensembles.}
\label{bk-quenched}
\end{center}
\end{table}
The results of this work (full QCD, $\beta = 5.7$,
$ 16^3 \times 40 $), Sharpe {\em et al.} 
\cite{sharpe20,kilcup30} 
(quenched QCD, $ \beta = 6.0 $, $ 24^3 \times 40 $),
Kilcup \cite{kilcup10,kilcup30} (quenched QCD, 
$ \beta = 6.0 $, $ 16^3 \times 40 $) and
Ukawa {\em et al.} \cite{japan0,japan20} 
(quenched QCD, $ \beta = 6.0 $,
$ 24^3 \times 40 $)
are compared in Table \ref{bk-quenched}.
From Table \ref{bk-quenched}, we conclude that
the dependence of $ B_K $ on $ m_{\rm sea} $
is too weak to detect in our numerical simulations
(in other words, the effect of quenched approximation
is less than 15 \% in $ B_K $ measurements).
This has been predicted 
by the quenched chiral perturbation theory
in the limit of $ m_s \rightarrow m_d $
\cite{sharpe1,sharpe30}.
\subsection{Fitting Procedures for $ B_K $}
$ B_K $ describes  $ K^0 -\bar{K}^0 $
mixing at the energy scale of about 500 MeV
({\em i.e.} in the low energy limit of QCD dynamics).
It is not known how to calculate
the dependence of $ B_K $ on the valence quark mass
directly from the QCD Lagrangian.
For this reason we adopt the chiral effective Lagrangian
(equivalent to current algebra), which is
valid in the energy region below the 
$ \rho $ meson mass.
This chiral effective lagrangian is not a cure-all
solution to the low energy dynamics of QCD.
However, it gives us a reasonable guide to understand the
leading chiral behavior of QCD.
We will now use the predictions of the chiral effective
Lagrangian to interpolate between our $ B_K $ results
to make a $ B_K $ prediction for physical quark masses.

The corrections from chiral perturbation theory
to $ {\cal M}_K $ in full QCD were calculated in Ref. 
\cite{sharpe30,sonoda0,donoghue,bardeen}.
The results of that calculation \cite{sharpe30}
($ m_u = m_d \ne m_s$) in full QCD were 
\begin{eqnarray}
B_K^{\rm full \ QCD} & = & 
B \left[ 1 + I_2(m_K) - \frac{1}{4}
\left( \frac{5m_d + 7 m_s}{m_d + m_s} \right) 
I_1(m_{\eta}) \right.
\nonumber \\
& &  \hspace*{10 mm} 
-  \left. \frac{1}{4}
\left(\frac{3m_d + m_s}{m_d + m_s} \right)
I_1(m_{\pi}) + O(m_K^4 \ln^2(m_K^2)) \right]
\nonumber \\
& = &
B \left[ 1 - ( 3 + \frac{1}{3} \epsilon^2)
\frac{m_K^2}{ (4\pi f_{\pi})^2 }
\ln\left( \frac{m_K^2}{ (4\pi f_{\pi})^2 } \right)
\right.
\nonumber \\
&  & \hspace*{10 mm}
+ \left. c_1 \frac{m_K^2}{ (4\pi f_{\pi})^2 }
+ c_2 \epsilon^2 \frac{m_K^2}{ (4\pi f_{\pi})^2 }
+ O( m_K^2 \epsilon^4) \right]
\label{eq-bk-full}
\end{eqnarray}
where
$ \epsilon $ is defined as
\begin{eqnarray}
\epsilon \equiv \left( 
\frac{m_s -m_d}{m_s + m_d} \right)
\ .
\end{eqnarray}
$ c_1 $, $c_2 $ are unknown coefficients,
but can be determined by numerical simulation
on the lattice.
The results from quenched chiral perturbation theory
\cite{sharpe30}
($ m_u = m_d \ne m_s$) are
\begin{eqnarray}
B_K^{\rm quenched} & = & 
B \left[ 1 + I_2(m_K)
- \frac{3 m_d + m_s}{2 m_d + 2 m_s} I_1(m_{dd})
\right.
\nonumber \\
& & 
\hspace*{10 mm}
- \frac{m_d + 3 m_s}{2 m_d + 2 m_s} I_1(m_{ss})
+ 
\delta \left\{ \frac{2 -\epsilon^2}{2 \epsilon} \ln
\left( \frac{1-\epsilon}{1+\epsilon} \right) + 2 \right\}
\nonumber \\
& & 
\hspace*{10 mm}
\left. 
+ O(m_K^4 \ln^2(m_K^2) ) \right]
\nonumber \\
& = &
B \left[ 1 - (3 + \epsilon^2 ) 
\frac{m_K^2}{ (4\pi f_{\pi})^2 }
\ln\left( \frac{m_K^2}{ (4\pi f_{\pi})^2 } \right)
\right.
\nonumber \\
& &
\hspace{10mm}
+  c'_1 \frac{m_K^2}{ (4\pi f_{\pi})^2 }
+ c'_2 \epsilon^2 \frac{m_K^2}{ (4\pi f_{\pi})^2 }
\nonumber \\
& & \hspace*{10 mm}
+ \left. 
\delta \left\{ \frac{2 -\epsilon^2}{2 \epsilon} \ln
\left( \frac{1-\epsilon}{1+\epsilon} \right) + 2 \right\}
+  O(m_K^4 \ln^2(m_K^2) ) \right]
\nonumber \\
\label{eq-bk-quenched}
\end{eqnarray}
where quenched chiral perturbation \cite{sharpe30} predicts 
\begin{eqnarray}
\delta \equiv \frac{ A m_0^2 }{ N (4 \pi f)^2 } \approx 0.2 \ .
\end{eqnarray}
$ c'_1 $ and $ c'_2 $ are unknown coefficients.
The term proportional to $ \delta $ is a contribution of 
$ \eta' $ loops appearing in quenched QCD,
which are absent in full QCD.
Because of this term, $ B_K^{\rm quenched} $
has a singular branch point at $ \epsilon = \pm 1 $.
Note, $ B_K^{\rm full \ QCD} $ does not
have any singular branch point.
Equations (\ref{eq-bk-full}) and
(\ref{eq-bk-quenched}) are the theoretical
predictions for $ B_K $ as a function of the
light quark masses.
Let us choose our fitting function on the basis of
the predictions of (quenched) chiral perturbation
theory in  Eq. (\ref{eq-bk-full}) and (\ref{eq-bk-quenched}).
Parts of our numerical simulation 
of $ B_K $ are classified in the category of
partially quenched QCD
($ N_f = 2 $ but $ m_{\rm sea} \ne m_{\rm valence} $).
We notice that the coefficients
of those terms proportional to 
$ \epsilon^2 m_K^2 \ln ( m_K^2) $
in Eqn. (\ref{eq-bk-full}) and
Eqn. (\ref{eq-bk-quenched}) are different
from each other, which implies that
the coefficient of these terms should be determined
by our numerical data. 
We have tried a linear fitting function
($ B_K(m_q) = \alpha_1 + \alpha_2 m_q a $)
even though the linear term is a next to
leading order correction from the chiral perturbation
(as can be seen in Eqn. (\ref{eq-bk-full}) and 
(\ref{eq-bk-quenched})).
The $ \chi^2 $/(d.o.f.) for the linear covariance fitting
on the jack-knifed ensembles of $ B_K $ data 
calculated in the two spin trace form 
with the cubic wall source method is 17.2(37).
This implies that the fitting is poor
and that we need an additional term
({\em e.g.} $ m_qa \ln(m_qa) $) to fit the data.
Hence, we choose the first fitting function
as
\begin{eqnarray}
B_K(m_q)  = \alpha_1 + \alpha_2 m_q a \ln (m_q a)
+ \alpha_3 m_q a \ . 
\label{first-fit}
\end{eqnarray}
where $  \alpha_i $ ($ i = 1,2,3$) are unknown
coefficients to be determined
and $ m_q a = \frac{1}{2} (m_s + m_d) $.
In this first fitting function,
we neglected the effect of non-degenerate quark mass
pairs ({\em i.e.} terms proportional to $ \epsilon^2 $
are neglected).
From the chiral perturbation theory 
Eqn. (\ref{eq-bk-full}) and (\ref{eq-bk-quenched}),
the predicted value of the ratio 
$ \alpha_2/\alpha_1 $ is
\begin{eqnarray}
\frac{\alpha_2}{\alpha_1} & = & (3 + \Delta) 
\frac{2 \mu a}{ (4\pi f_\pi a)^2 } \ ,
\end{eqnarray}
where $ 0 \le \Delta \le 1 $.
Using our lattice QCD simulation results for $ \mu $
($ \mu \cong 2.3 a^{-1} $),
the ratio $ \alpha_2/\alpha_1 $ is written
in terms of poorly known quantity $ \Delta $:
\begin{eqnarray}
\frac{\alpha_2}{\alpha_1} \cong 40 + 13 \Delta \ .
\end{eqnarray}
Using the value of $ K $ meson mass and
strange quark mass in the particle data book
($ m_s \cong 100 \sim 300 MeV $ and
$ m_s/m_d \cong 25 $),
one can also obtain the ratio $ \alpha_2/\alpha_1 $
\begin{eqnarray}
\frac{\alpha_2}{\alpha_1} \cong (7.0 \sim 23 )  + (2.3 \sim 7.6) \Delta,
\end{eqnarray}
in terms of a poorly known quantity $ \Delta $, where
the range of values is chosen to reflect the large
uncertainty in the chiral perturbation theory prediction.
The $ B_K $ results 
of the above 3 parameter covariance fitting 
on the bootstrap ensembles 
are summarized in Table \ref{table_bk_3para} (unrenormalized)
and \ref{table_renorm_bk_3para} (tadpole-improved renormalized).
\begin{table}[p]

\centering

\begin{tabular}{|| c || c | c | c | c || c ||}
\hline
Trace Form (Source) & $ \alpha_1 $ & $ \alpha_2 $ & 
$ \alpha_3 $ & $ \chi^2 $/d.o.f. &  $ B_K (m_q) $ 
\\ \hline \hline
2TR (Cubic) & 0.336(65) & -11.2(33) & -24.1(93) & 1.05 & 6.600(21)
\\ \hline 
2TR (Even-Odd) & 0.346(89) & -10.1(39) & -21.9(112) & 1.87 & 6.375(29)
\\ \hline
1TR (Even-Odd) & 0.477(119) & -9.35(482) & -21.9(137) & 0.47 & 7.270(30)
\\ \hline
\end{tabular}

\caption{ Covariance fitting of unrenormalized $ B_K $ 
with 3 parameters: 
Here 2TR and 1TR represent
the two spin trace form and the one spin trace from respectively.
Cubic and Even-Odd imply the cubic wall source and the even-odd
wall source respectively and
$ B_K (m_q) $ means interpolation to the physical quark mass. }

\label{table_bk_3para}

\vspace*{5mm}

\begin{tabular}{|| c || c | c | c | c || c ||}
\hline
Trace Form (Source) & $ \alpha_1 $ & $ \alpha_2 $ & 
$ \alpha_3 $ & $ \chi^2 $/d.o.f. &  $ B_K (m_q) $ 
\\ \hline \hline
2TR (Cubic) & 0.372(62) & -9.83(32) & -21.6(95) & 1.06 & 0.655(21)
\\ \hline 
2TR (Even-Odd) & 0.416(86) & -7.43(37) & -15.9(102) & 2.03 & 0.636(28)
\\ \hline
1TR (Even-Odd) & 0.408(112) & -8.30(480) & -19.2(122) & 0.52 & 0.633(27)
\\ \hline
\end{tabular}

\caption{ Covariance fitting of tadpole-improved 
renormalized $ B_K $ with 3 parameters at the scale
of $ \mu = \pi/a $: 
Here 2TR and 1TR represent
the two spin trace form and the one spin trace from respectively.
Cubic and Even-Odd imply the cubic wall source and the even-odd
wall source respectively and
$ B_K (m_q) $ means interpolation to the physical quark mass. }

\label{table_renorm_bk_3para}

\end{table}
From Table \ref{table_bk_3para} and \ref{table_renorm_bk_3para},
we observe that the renormalization with tadpole improvements
makes $ B_K (m_q) $ in agreement between the one spin trace form
and the two spin trace form once they are obtained with the same source
method and covariance-fitted in the same range.
The covariance fitting results for the ratio $ \alpha_2/\alpha_1 $
were not consistent among the various measurements,
mainly because the 3 parameters have 
a wide domain to fit the 7 $ B_K $ data points 
as a function of the average quark mass.
This gave us a motivation for the second fitting trial function,
which will have only two parameters.

Our second fitting trial function is chosen such that
the ratio $ \alpha_1 / \alpha_2 $ is fixed to the
case of the degenerate quark anti-quark pair
($ \epsilon = 0 $). 
The reason is that this ratio is universal for the degenerate
quark pair regardless of quenched approximation and that
our $ \epsilon $ values are small enough to take into consideration
later as a perturbative expansion parameter.
The second fitting trial function is
\begin{eqnarray}
B_K(m_K)  = \alpha_1 
\left[ 1 - 
3 \frac{m_K^2}{ (4\pi f_{\pi})^2 }
\ln\left( \frac{m_K^2}{ (4\pi f_{\pi})^2 } \right)
\right]
+ \alpha_2 \frac{m_K^2}{ (4\pi f_{\pi})^2 }  
\label{second-fit} \ ,
\end{eqnarray}
where $ \alpha_i $ ($i = 1,2$) are unknown coefficients.
We take the effective mass ($ m_K $) of $ f_K $
($ \langle 0 \mid  A_{\mu} \mid K^0 \rangle $)
measurements in Table \ref{bk-table} as a definition of $ m_K $ in
Eq. (\ref{second-fit}).
In this second fitting trial function, we set
the coefficient of the leading term in the chiral 
perturbation expansion to the theoretically expected 
value in Eq. (\ref{eq-bk-full}, \ref{eq-bk-quenched})
and also we again neglect the effect of
non-degenerate quark mass pairs.
When we set $ a^{-1} $ = 1.8 GeV from the
$ \rho $ meson mass and choose $ f_{\pi} $ = 93 MeV,
the covariance fitting results 
for the {\em two spin trace form (cubic wall source)}
are summarized in Table \ref{table_bk_2para} and 
\ref{table_renorm_bk_2para}.
\begin{table}[p]

\centering

\begin{tabular}{|| c || c | c | c || c ||}
\hline
Trace Form (Source) & $ \alpha_1 $ & $ \alpha_2 $ & 
$ \chi^2 $/d.o.f. &  $ B_K (m_K) $ 
\\ \hline \hline
2TR (Cubic) & 0.293(14) & 0.397(66) & 0.897 & 0.638(21)
\\ \hline 
2TR (Even-Odd) & 0.289(21) & 0.334(82) & 1.53 & 0.619(28)
\\ \hline
1TR (Even-Odd) & 0.346(24) & 0.191(84) & 0.70 & 0.702(35)
\\ \hline
\end{tabular}

\caption{ Covariance fitting of unrenormalized $ B_K $ 
with 2 parameters: 
Here 2TR and 1TR represent
the two spin trace form and the one spin trace from respectively.
Cubic and Even-Odd imply the cubic wall source and the even-odd
wall source respectively and
$ B_K (m_K) $ means interpolation to the physical K meson mass.}

\label{table_bk_2para}

\vspace*{5mm}

\begin{tabular}{|| c || c | c | c || c ||}
\hline
Trace Form (Source) & $ \alpha_1 $ & $ \alpha_2 $ & 
$ \chi^2 $/d.o.f. &  $ B_K (m_K) $ 
\\ \hline \hline
2TR (Cubic) & 0.301(14) & 0.300(67) & 0.89 & 0.636(21)
\\ \hline 
2TR (Even-Odd) & 0.297(21) & 0.255(83) & 1.83 & 0.619(28)
\\ \hline
1TR (Even-Odd) & 0.300(21) & 0.184(81) & 0.47 & 0.612(29)
\\ \hline
\end{tabular}

\caption{ Covariance fitting 
of tadpole-improved renormalized $ B_K $ with 2 parameters
at the scale of $ \mu = \pi/a $: 
Here 2TR and 1TR represent
the two spin trace form and the one spin trace from respectively.
Cubic and Even-Odd imply the cubic wall source and the even-odd
wall source respectively and
$ B_K (m_K) $ means interpolation to the physical K meson mass.}

\label{table_renorm_bk_2para}

\end{table}
From Table \ref{table_bk_2para} and \ref{table_renorm_bk_2para},
we notice that the fitting results for
the tadpole-improved renormalized $ B_K $ 
in various types of the measurements
are in good agreement with one another, while the fitting results
for the unrenormalized $ B_K $ are not consistent between the
one spin trace form and the two spin trace form.

\begin{figure}[t]

\centering

\epsfxsize=\hsize\epsfbox{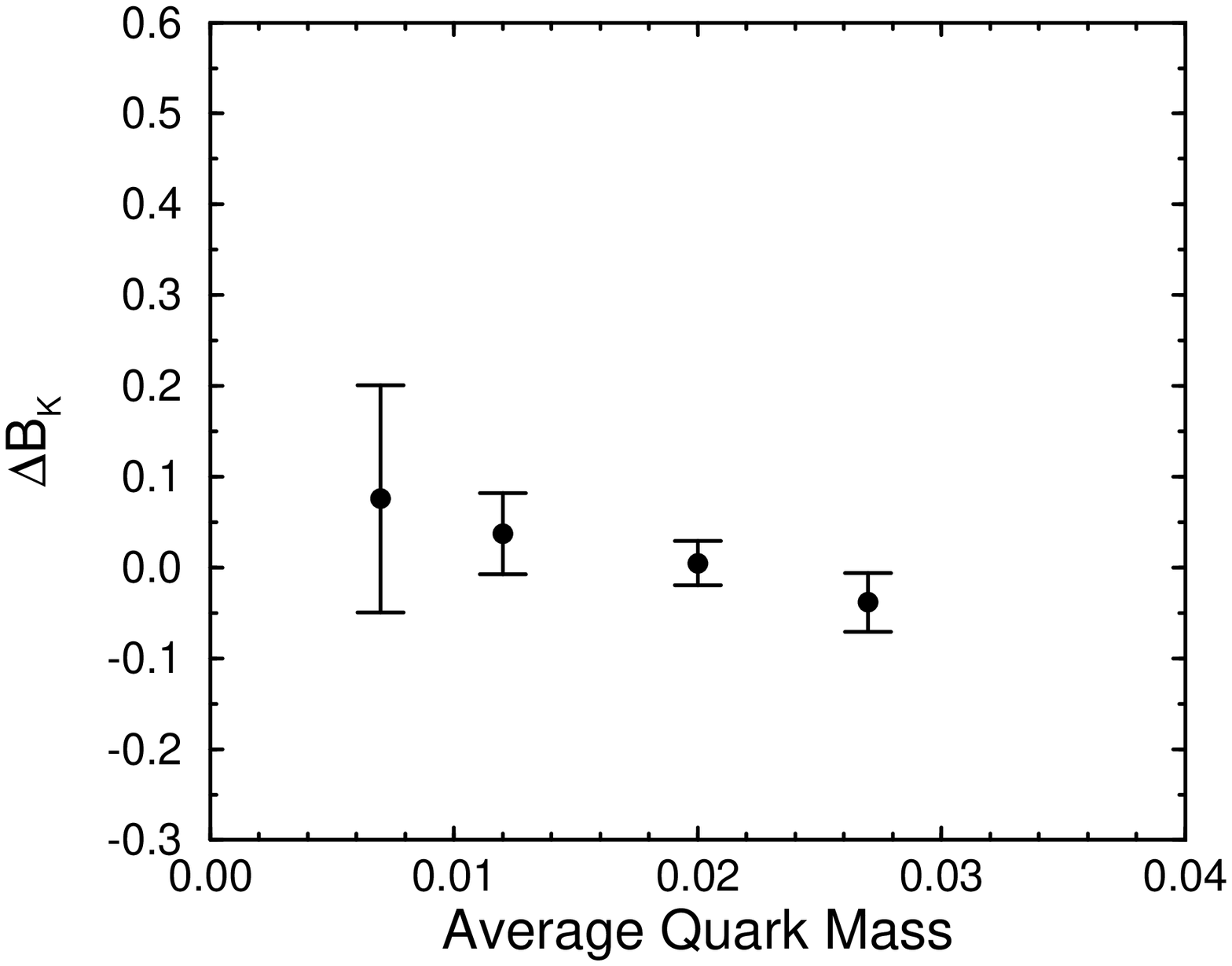}

\caption{$ \Delta B_K $ with respect to average quark mass.
The data are obtained using
the two spin trace formalism with the cubic wall
source method.}

\label{fig:19}

\end{figure}

\begin{figure}[t]
 
\centering

\epsfxsize=\hsize\epsfbox{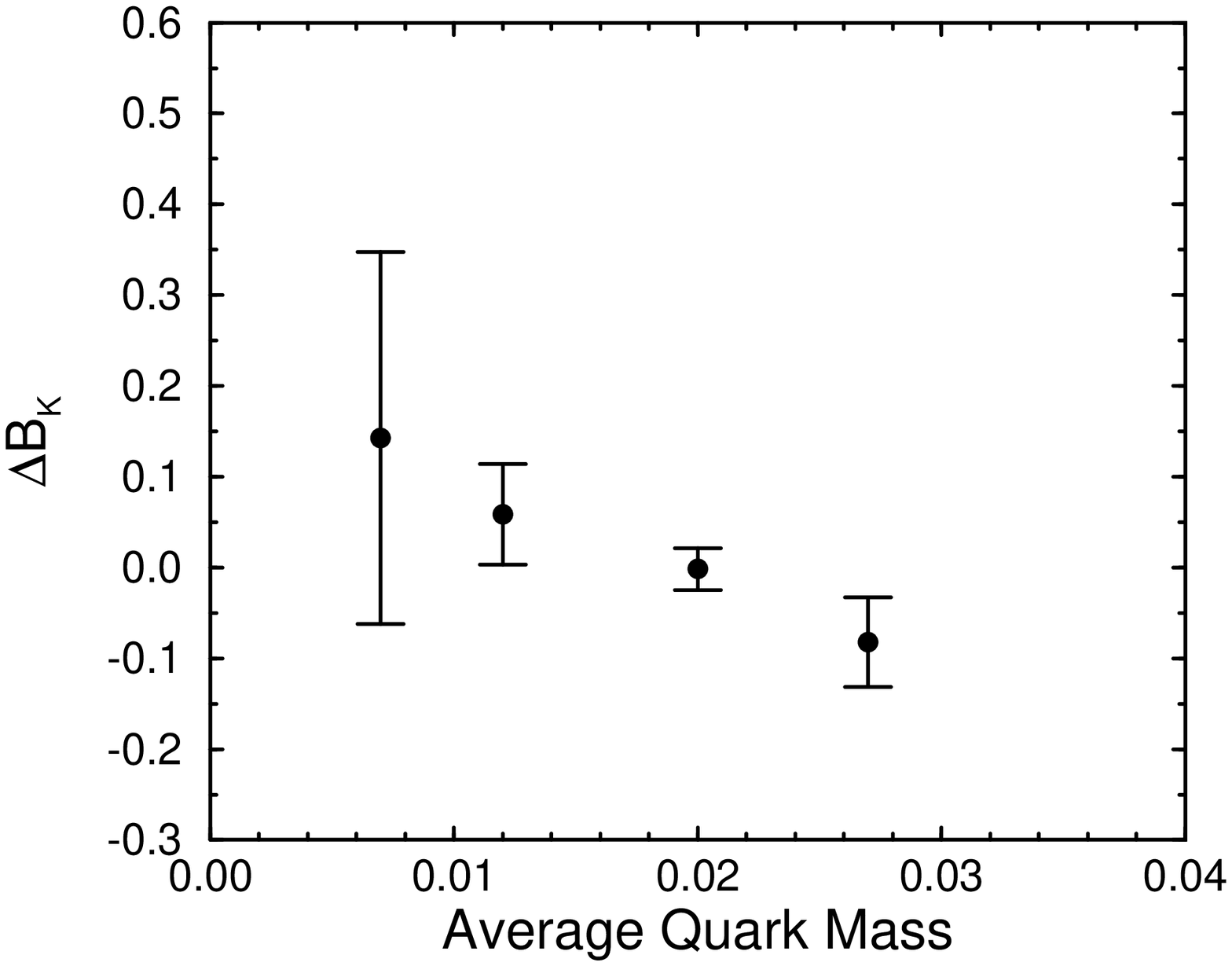}

\caption{$ \Delta B_K $ with respect to average quark mass.
The data are obtained using
the two spin trace formalism with the even-odd wall
source method.}
 
\label{fig:20}

\end{figure}

\begin{figure}[t]
 
\centering

\epsfxsize=\hsize\epsfbox{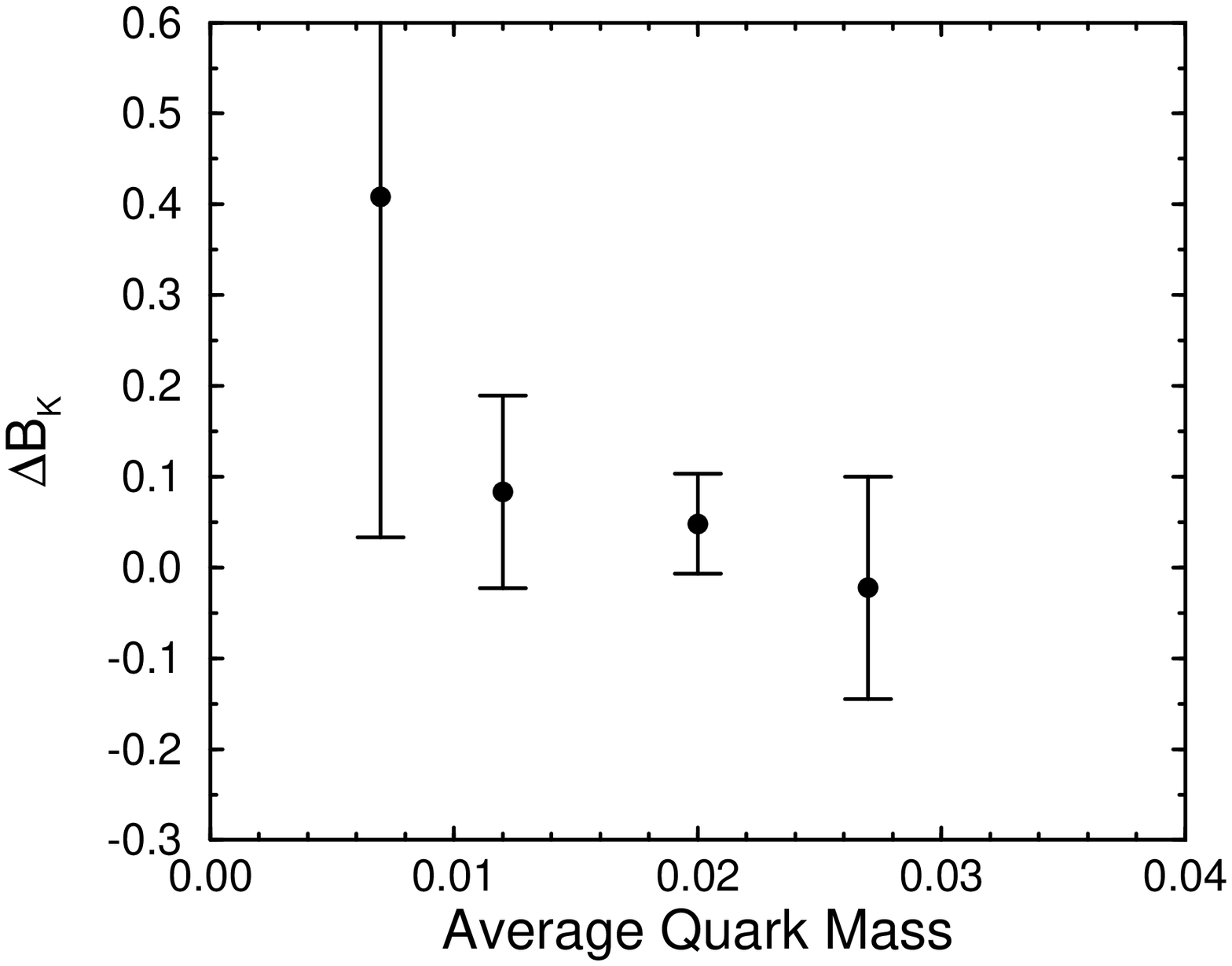}
 
\caption{$ \Delta B_K $ with respect to average quark mass.
The data are obtained using
the one spin trace formalism with the even-odd wall
source method.}
 
\label{fig:21}

\end{figure}

Let us discuss how we can detect the effect of
non-degenerate quark anti-quark pairs on $ B_K $ 
({\em i.e.} the dependence of $ B_K $ on $ \epsilon $).
The strategy is the following.
First, we divide the numerical data in two parts:
one part (we call this the {\em degenerate part})
contains only the numerical data  for
degenerate quark mass pairs \{(0.01, 0.01),
(0.02, 0.02), (0.03, 0.03)\} and the other part
(the {\em non-degenerate part})
contains only the data for non-degenerate quark mass
pairs \{(0.004, 0.01), (0.004, 0.02), (0.004, 0.05),
(0.01, 0.03)\}.
Next, we fit the {\em degenerate part} to
the second trial function in Eqn. (\ref{second-fit})
which are supposed to be exact for the degenerate
quark mass pairs up to the given order
in the chiral perturbative expansion.
For the {\em degenerate part},
\begin{eqnarray}
B_K^{\rm degenerate}(m_K) \equiv \alpha_1 
\left[ 1 - 
3 \frac{m_K^2}{ (4\pi f_{\pi})^2 }
\ln\left( \frac{m_K^2}{ (4\pi f_{\pi})^2 } \right)
\right]
+ \alpha_2 \frac{m_K^2}{ (4\pi f_{\pi})^2 }  
\label{degenerate-fit}
\end{eqnarray}
Now we define a function
which represents the effect of non-degenerate
quark mass pairs:
\begin{eqnarray}
\Delta B_K(m_K) \equiv 
\frac{ B_K(m_K, \epsilon) - B_K^{\rm degenerate}(m_K)}
{\epsilon^2}
\end{eqnarray}
where $ B_K(m_K, \epsilon) $ are our numerical
data with non-vanishing
$ \epsilon^2 = (m_s-m_d)^2/(m_s+m_d)^2 $ and
$  B_K^{\rm degenerate}(m_K) $ represents the
fitting function for the degenerate part,
given in Eq. (\ref{degenerate-fit}).
Finally, we try to find a functional form, if possible, to
fit $ \Delta B_K(m_K) $ numerical data to.
In Figure \ref{fig:19} (two spin trace form, cubic wall source),
Figure \ref{fig:20} (two spin trace form, even-odd wall source)
and Figure \ref{fig:21} (one spin trace form, even-odd wall source),
we plot $ \Delta B_K $ with respect to the
average quark mass, only for the non-degenerate
quark anti-quark pairs.
As you can see in the figures, it is hard to find a
functional form which can explain all of the data.
From these figures, we notice that the dependence of
$ B_K $ on $ \epsilon^2 $ is extremely small
(especially in the domain
near the physical $ K $ meson mass). 
Hence, we conclude that
we could not detect any significant effect of
non-degenerate quark mass pairs on $ B_K $ within the 
precision of our numerical study.

As a conclusion to this section, let us 
present our best value of $ B_K $.
Since the data of the cubic wall source method
has better statistics than that of
the even-odd wall source method,
we shall quote the covariance
fitting results of the cubic wall source data
as our best value.
For the second trial function, the uncertainty
in the lattice scale $ a $ and the decay constant $ f_\pi $ ($ f_K $)
produces uncertainty
in $ B_K (m_K) $ of the same order of magnitude as the statistical
uncertainty, while covariance fitting to 
the first trial function is not as much sensitive to the lattice
scale $ a $ and the decay constant $ f_\pi $.
Hence, we choose the results of the covariance fitting
to the first fitting trial function with three parameters
on the bootstrap ensembles
as our best value.
Interpolated from the first fitting trial function,
the physical results are
\begin{eqnarray}
\mbox{unrenormalized } B_K (m_K) & = & 0.660(21) \nonumber \\ 
\mbox{renormalized (N.D.R.) } 
B_K (m_K, \mu=\frac{\pi}{a}) & = & 0.655(21) \ ,
\end{eqnarray}
where N.D.R. implies {\em naive dimensional regularization} 
scheme and the errors represent purely statistical uncertainty.
Here the physical results imply the $ B_K $ values for 
the physical kaon mass ($ m_K = 497.7 $ MeV).
Here, we have completely neglected
the systematic errors related to 
the scale ($ a $) uncertainty,
the coupling ($ g^2_{\overline{\rm MS}} $) uncertainty,
the contamination from
the unwanted operator mixing,
and 
the contamination from unwanted hadronic
states which can couple to the operators used.
We also could not control
finite volume effects, finite temperature effects,
or finite lattice spacing effects.
\section{Conclusion}
Here, we summarize what we have learned through
the numerical simulation of $ B_K $ and what needs
further investigation in the future.
The results for  $ B_K $ from the 
improved wall source (cubic wall source)
were in good agreement with those of the conventional
even-odd wall source. 
It is shown that the cubic wall source suppresses the
contamination from the wrong flavor channels efficiently.
However, the cubic wall source takes 
four times more computational time than the
even-odd wall source.
In the limit of $ a \rightarrow 0 $,
the SU(4) flavor symmetry is supposed to
be recovered and so
there will be a serious contamination
from pseudo Goldstone pions 
with various flavor structures and the various
$ \rho $ mesons. 
The cubic wall source is
quite promising in the
weak coupling region
to exclude the contamination
from unwanted hadronic eigenstates.
We can transcribe the continuum $ \Delta S = 2 $
operator to the lattice with staggered fermions 
in two different ways.
Theoretically, both formalisms
of operator transcription
must be equivalent to each other in the
limit $ a \rightarrow 0 $.
The numerical results in the one spin trace formalism
were consistent with those in the two spin trace formalism
after the proper renormalization 
(with either tadpole or RG ($ \overline{MS} $) improvement).
We have learned how important the proper renormalization is,
as well as the careful choice of the coupling constant for the
perturbative expansion.
We have tried to understand the effects of 
the non-degenerate quark masses on $ B_K $ and 
the individual components making up $ B_K $.
The  effects of the non-degenerate quark antiquark pairs
on $ B_K $ were too small to observe 
within the precision of our numerical simulation
(especially near the region of physical $ K $ meson mass).
Why this effect is so small needs more careful
theoretical investigation.
Chiral perturbation theory suggests that 
$ B_{A1} $ (axial part of $ B_K $ with one color loop)
is the best observable to
detect the enhanced chiral logarithms
which are not expected to be a function of  
quark mass difference.
We observed an additional divergence
which depends on the
quark mass difference.
This additional divergence needs more
thorough investigation and understanding.
We wonder whether
partially quenched chiral perturbation
can explain this additional divergence, or
how much finite volume effects on
$ B_{A1} $ depend on the lightest mass of the non-degenerate
quark anti-quark pair. 
Qualitatively,  chiral perturbation
theory is consistent with our
numerical work.
However, 
more theoretical research on the (partially)
quenched chiral perturbation and its quenched chiral
logarithms is necessary to see whether the
(partially) quenched chiral perturbation can explain
the effects of the non-degenerate quark antiquark pairs
on $ B_K $ and its individual components. 
We could not observe any dynamical fermion effect
on $ B_K $.
It is  difficult to understand why
internal fermion loops are so unimportant for $ B_K $,
since the Dirac eigenvalue spectrum of 
quarks is supposed to be completely
different between quenched QCD and full QCD.
This also needs further theoretical understanding.
Through the numerical study of $ B_K $ in this paper,
we have learned that the cubic source method is
promising for the weak coupling simulation and
that the one spin trace formalism is consistent with
the two spin trace formalism.
It is true that lattice QCD results for $ B_K $ 
are more solid and believable after this work.
However, it is also true that
there are many details which need more
thorough investigation and understanding.  
\section{Acknowledgement}
One of the authors (W. Lee) is indebted a lot to Norman H. Christ.
This work could not have been done without his consistent help
and encouragement. 
Helpful discussion with Robert D. Mawhinney
at the early stage of this work is acknowledged with
gratitude.
One of the authors (W. Lee) would like to express his sincere
gratitude to Donald Weingarten at IBM T.J. Watson Research Center
for his superb lecture of bootstrap analysis.
One of the authors (W. Lee) would like to thank Gregory Kilcup
for his kind help in many ways during his visit to Columbia University.
Helpful discussion with Stephen Sharpe and
A. Ukawa during Santa Fe workshop is acknowledged with sincere
gratitude.
We would like to thank Zhihua Dong for his kind help in 
gauge fixing programming.
Helpful conversation with Shailesh Chandrasekaharan, Dong Chen, and
Decai Zhu is acknowledged with gratitude.
\clearpage

\input epsf

%
%

%
%

%
%

%
%

%
%

%
%

%
%

%
%

%
%

%
%

\clearpage

%
%

%
%

%
%

%
%

%
%

%
%

%
%

%
%

%
%

%
%

\clearpage

%
%
%
%
%
\end{document}